%% file: excursion_set.tex
\documentclass[a4paper,11pt]{article}
\pdfoutput=1 
\usepackage{jcappub}

\usepackage{amsmath}
\usepackage{bm}
\usepackage{graphicx}
\usepackage{enumitem}
\usepackage{geometry}
\usepackage{xspace}
\usepackage{pdflscape}
\usepackage{placeins}
\usepackage{epstopdf}
\usepackage{comment}
\usepackage{subcaption}
\usepackage{tensor}
\usepackage{multirow}

\makeatletter
\gdef\@fpheader{}
\g@addto@macro\bfseries{\boldmath}
\makeatother

\input{newcommands}

\newcommand{\mPBH}{M}
\newcommand{\OmegaPBH}{\Omega_{\scriptscriptstyle{\mathrm{PBH}}}}

\subheader{}

\title{Primordial black holes from metric preheating: mass fraction in the excursion-set approach}

\author{Pierre Auclair,}
\author{Vincent Vennin} 

\affiliation{Laboratoire Astroparticule et Cosmologie, CNRS Universit\'e de Paris, 75013 Paris, France}

\emailAdd{pierre.auclair@apc.in2p3.fr}
\emailAdd{vincent.vennin@apc.univ-paris7.fr}

\date{today}

\begin{document}

\sloppy

\abstract{
We calculate the mass distribution of Primordial Black Holes (PBHs) produced during metric preheating. After inflation, the oscillations of the inflaton at the bottom of its potential source a parametric resonant instability for small-scale scalar perturbations, that may collapse into black holes. After reviewing in a pedagogical way different techniques that have been developed in the literature to compute mass distributions of PBHs, we focus on the excursion-set approach. We derive a Volterra integral equation that is free of a singularity sometimes encountered, and apply it to the case of metric preheating. We find that if the energy density at which the instability stops, $\rho_\Gamma$, is sufficiently smaller than the one at which inflation ends, $\rho_\uend$, namely if $\rho_\Gamma^{1/4}/\rho_\uend^{1/4}< 10^{-5}(\rho_\uend^{1/4}/10^{16}\mathrm{GeV})^{3/2}$, then PBHs dominate the universe content at the end of the oscillatory phase. This confirms the previous analysis of \Refa{Martin:2019nuw}. By properly accounting for the ``cloud-in-cloud'' mechanism, we find that the mass distribution is more suppressed at low masses than previously thought, and peaks several orders of magnitude above the Hubble mass at the end of inflation. The peak mass ranges from $10$ g to stellar masses, giving rise to different possible cosmological effects that we discuss. 
}



\maketitle

\section{Introduction}
\label{sec:intro}
Since Primordial Black Holes (PBHs) were proposed almost 50 years ago~\cite{Carr:1974nx,1975ApJ...201....1C}, it has been realised that they can be relevant in various aspects of cosmology, ranging from dark matter~\cite{Chapline:1975ojl} and the generation of large-scale structures through Poisson fluctuations~\cite{Meszaros:1975ef,Afshordi:2003zb} to the seeding of supermassive black holes in galactic nuclei~\cite{Carr:1984id, Bean:2002kx}. More recently, they have attracted even more attention as it was pointed out that they may account for the progenitors of the black-hole merging events detected by the LIGO/Virgo collaboration~\cite{LIGOScientific:2018mvr} through their gravitational wave emission, see \eg \Refs{Sasaki:2016jop, Abbott:2020mjq}.  There are several observational bounds that constrain the abundance of PBHs in various mass ranges (for a recent review, see \eg \Refa{Carr:2020gox}).

PBHs are expected to form when large density fluctuations re-enter the Hubble radius and collapse into black holes. Their abundance is usually computed by assuming that they are rare objects that are formed at around a single scale, and the probability that a given region of the universe ends up in a black hole can be inferred from the knowledge of the primordial curvature power spectrum at that scale. Such an approach may however fail in cases where PBHs are abundantly produced, and/or if they arise over a wide range of masses. This could notably be the case for PBHs with masses smaller than $10^9\,\mathrm{g}$, which Hawking evaporate before big-bang nucleosynthesis and are therefore not limited by observational constraints.

A prototypical example of a mechanism leading to such ultra-light, yet extremely abundant, PBHs, is the parametric instability of single-field metric preheating~\cite{Jedamzik:2010dq, Easther:2010mr} (see \Refa{Green:2000he} for multiple-field setups). After inflation, the oscillations of the inflaton at the bottom of its potential source a parametric instability in the equation of motion of scalar perturbations, that are enhanced on small scales. In \Refa{Martin:2019nuw}, it was shown that the production of ultra-light PBHs from this instability  is so efficient that they can quickly come to dominate the universe content, such that reheating no longer occurs because of the inflaton decay, but rather through PBHs evaporation.

Although these conclusions lead to a substantial change in the cosmological scenario, they were however reached by employing the usual estimate for PBH abundance, whose usage is questionable in contexts in which PBHs are abundant. The goal of this paper is to re-examine this calculation, in the light of more refined techniques that were originally proposed for large-scale structures but that can (and have) also be applied to PBHs. This will allow us to investigate generic properties of the expected mass distribution of ultra-light black holes, in regimes in which they densely populate the primordial universe.

This could have important consequences for various physical phenomena associated to those black holes. For instance, it was recently shown~\cite{Papanikolaou:2020qtd} that gravitational waves induced at second order by the gravitational potential underlain by ultra-light PBHs lead to a stochastic background that might be detected in future gravitational-wave experiments, and that is even already excluded in some regimes. Since the amplitude and frequency coverage of this background strongly depend on the details of the mass distribution of PBHs, it seems important to derive robust predictions for the scenarios in which they are produced.

The rest of this paper is organised as follows. In \Sec{sec:MetricPreheating}, we briefly describe the mechanism of metric preheating and the production of PBHs that is associated to it. In \Sec{sec:MassFraction}, we review the different techniques that have been proposed to compute the abundance of objects formed from gravitational collapse. Our main goal is to identify those that are best suited to the problem at hand, but we also designed this section as a pedagogical introduction to the calculation of the mass fraction, trying to highlight some aspects that are often left implicit. This section may however be skipped by readers already familiar with the topic. In \Sec{sec:PBH}, we apply one of the methods introduced in \Sec{sec:MassFraction}, namely the excursion-set approach, to the calculation of the mass fraction of PBHs arising from metric preheating. We finally present our conclusions in \Sec{sec:Discussion}, and the paper ends with several appendices where various technical aspects of the calculation are deferred.
\section{Metric preheating}
\label{sec:MetricPreheating}
In this section, we briefly review the physics of metric preheating. More details can be found in \Refs{Jedamzik:2010dq, Jedamzik:2010hq, Martin:2019nuw, Martin:2020fgl}. If a homogeneous and isotropic universe, described by the Friedmann-Lema\^itre-Robertson-Walker metric $\dd s^2 = -\dd t^2 + a^2(t) \dd x^2$ where $a$ is the scale factor, is dominated by a single scalar field $\phi$, scalar perturbations are described by a single gauge-invariant combination of fluctuations in the scalar field and in the metric components, the so-called Mukhanov-Sasaki variable~\cite{Mukhanov:1981xt,Kodama:1985bj}. Its equation of motion in Fourier space is given by~\cite{Mukhanov:1990me}
\begin{equation}
	\begin{aligned}
  \label{eq:eomv}
  v_{\bm k}''+\left[k^2-\frac{\left(a\sqrt{\epsilon_1}\right)''}{a\sqrt{\epsilon_1}}\right]v_{\bm k} = 0\, .
	\end{aligned}
\end{equation}
In this expression, a prime denotes derivative with respect to conformal time $\eta$ (related to cosmic time via $\dd t = a \dd \eta$), and $\epsilon_1 = -\dot{H}/H^2$ is the first slow-roll parameter, where $H=\dot{a}/a$ is the Hubble parameter. If the inflaton $\phi$ oscillates at the bottom of a quadratic potential $V(\phi)=m^2\phi^2/2$,\footnote{As the amplitude of the oscillations get damped, the leading order in a Taylor expansion of the function $V(\phi)$ around its minimum quickly dominates, which yields a quadratic potential unless there is an exact cancellation at that order.} the scale factor undergoes oscillations too (superimposed to an average matter-like behaviour), and \Eq{eq:eomv} can be put in the form~\cite{Jedamzik:2010dq}
\begin{equation}
	\begin{aligned}
\label{eq:Mathieu:v}
  \frac{\dd^2}{\dd z^2}\left(\sqrt{a} v_{\bm k}\right)
  +\left[A_{\bm k}-2q\cos(2z)\right]\left(\sqrt{a} v_{\bm k}\right)=0\, ,
	\end{aligned}
\end{equation}
with
\begin{equation}
	\begin{aligned}
\label{eq:A:q:def:metric:preheating}
  A_{\bm k}=1+\frac{k^2}{m^2a^2}, \quad
  q=\frac{\sqrt{6}}{2}\frac{\phi_\uend}{\Mp}\left(\frac{a_\uend}{a}\right)^{3/2}\, .
	\end{aligned}
\end{equation}
In those expressions, $a_\uend$ and $\phi_\uend$ are the values of $a$ and $\phi$ at the onset of the oscillating phase, \ie at the end of inflation, $\Mp$ is the reduced Planck mass, and $z\equiv mt+\pi/4$.

If $A_{\bm k}$ and $q$ were constant, this equation would be of the Mathieu type, and it would feature parametric resonant instabilities when $A_{\bm k}$ and $q$ fall into the instability bands. In  \Refa{Jedamzik:2010dq} (see also \Refa{Martin:2020fgl} where the perturbative decay of the inflaton is included in the analysis), the time dependence of $A_{\bm k}$ and $q$ is shown to be sufficiently slow to be considered as adiabatic, and the resonant instability takes place when $A_{\bm k}$ and $q$ cross the instability bands. At the end of inflation, the displacement of the field away from the minimum of its potential is typically of order the Planck mass, so \Eq{eq:A:q:def:metric:preheating} indicates that $q$ starts out being of order one and quickly decreases afterwards. This means that one falls into the regime of ``narrow resonance'', $q\ll 1$, in which the boundaries of the first instability
band are given by $1-q < A_{\bm k} < 1+q$, which here translates into
\begin{align}
\label{eq:instability:band:1}
k < a \sqrt{3 H m}.
\end{align}
Since the universe behaves as matter-dominated during the oscillatory phase, $a\sqrt{H} \propto a^{1/4}$, the upper bound~\eqref{eq:instability:band:1} increases with time, and the range of modes subject to the instability widens as time proceeds. Inside the first instability band, the Floquet index of the unstable mode is given by $\mu_{\bm{k}} \simeq q/2$, so $v_{\bm{k}}\propto a^{-1/2} \exp(\int \mu_{\bm{k}} \dd z)\propto a$~\cite{Finelli:1998bu,Jedamzik:2010dq}.

\begin{figure}[t]
\begin{center}
  \includegraphics[width=1.0\textwidth]{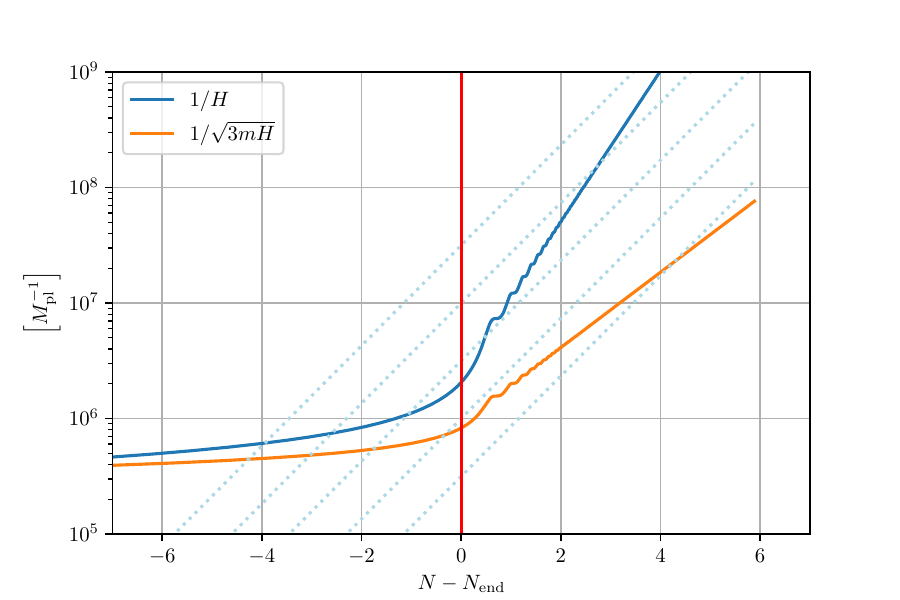}
\caption{Evolution of the physical scales appearing in \Eq{eq:instability:band:2}, with time parametrised by the number of \efolds~$N=\ln a$ (counted from the end of inflation). The blue line
  represents the Hubble radius $1/H$, the orange line the new
  length scale $1/\sqrt{3Hm}$ and the dotted lines the physical
  wavelengths of modes of interest, which may enter the instability band after inflation, during the oscillatory phase.
  Here the Klein-Gordon equation for the inflaton field has been solved for the quadratic potential $V(\phi) = m^2 \phi^2/2$, where $m = 10^{-6} \Mp$.
}
\label{fig:scale}
\end{center}
\end{figure}

Note that during the oscillatory phase, $H\ll m$, so the upper bound in \Eq{eq:instability:band:1} corresponds to a sub-Hubble scale. As a consequence, the modes subject to the instability are (i) all super-Hubble modes and (ii) those sub-Hubble modes such that $aH<k<a\sqrt{3 H m}$. For super-Hubble modes, the fact that $v\propto a$ implies that the curvature perturbation is simply conserved, which is a well-know result~\cite{Mukhanov:1981xt, Kodama:1985bj}, and the dynamics of super-Hubble scales is therefore not affected by the oscillations. For sub-Hubble scales however, in a matter-dominated background, the overdensity $\delta =\delta\rho/\bar{\rho}$ (where $\bar{\rho}$ is the energy density of the background) is related to the Mukhanov-Sasaki variable via $\delta_{\bm{k}}\propto [k/(aH)]^2 v_{\bm{k}}/(a \Mp)$ (notice that, at sub-Hubble scales, there is no gauge ambiguity in the definition of the density contrast, see \Sec{sec:Remove:Super:Horizon} for further details), so the fact $v_{\bm{k}} \propto a$ implies that
\begin{equation}
	\begin{aligned}
\label{eq:delta:a}
\delta_{\bm{k}}\propto a\, ,
	\end{aligned}
\end{equation}
and the density contrast grows inside the band
\begin{equation}
	\begin{aligned}
\label{eq:instability:band:2}
aH<k<a\sqrt{3 H m}.
	\end{aligned}
\end{equation}
The scales appearing in this relation are displayed in
\Fig{fig:scale}. An instability is triggered if the physical
wavelength of a mode (dotted line) is smaller than the Hubble radius
(blue line) during the oscillatory phase and larger than the new
scale $1/\sqrt{3Hm}$ (orange line). Let us note that this  implies that the instability
only concerns modes that are inside the Hubble radius at the end of
the oscillatory phase, which is not the case for the scales probed in
the Cosmic Microwave Background (CMB).

The behaviour~\eqref{eq:delta:a} signals that inside the resonance band, scalar-field inhomogeneities behave as pressureless matter fluctuations in a pressureless matter universe. As a consequence, an over-density $\delta_R$ over the length scale $R$ eventually collapses into a PBH, and in \Refa{Goncalves:2000nz} (see also appendices A and B of \Refa{Martin:2019nuw}), it is shown that this occurs after a time
\begin{equation}
	\begin{aligned}
\label{eq:t:coll}
\Delta t_{\mathrm{collapse}}
= \frac{\pi}{H\left[t_{\mathrm{bc}}(R)\right]
  \delta_R^{3/2}\left[t_{\mathrm{bc}}(R)\right]},
	\end{aligned}
\end{equation}
where $t_{\mathrm{bc}}(R)$ denotes the ``band-crossing'' time, \ie the time at which the scale $R$ crosses in the instability band~(\ref{eq:instability:band:2}). Assuming that the instability ends when the Hubble scale reaches a certain value that we denote $H_\Gamma$,\footnote{This could correspond for instance to the time when $H$ drops below $\Gamma$, the decay rate of the inflaton into other degrees of freedom, in the context of perturbative reheating~\cite{Albrecht:1982mp,Abbott:1982hn,Kofman:1997yn} (hence the notation).} at a time $t_\Gamma$, a black hole forms if $t_\uend + \Delta t_{\mathrm{collapse}} < t_\Gamma$, which leads to a lower bound on the density contrast, namely
\begin{equation}
	\begin{aligned}
\label{eq:deltac:bc}
\delta_R[t_{\mathrm{bc}}(R)] >  \left(\frac{3\pi}{2}\right)^{2/3}\left[\frac{H_{\mathrm{bc}}(R)}{H_\Gamma}-1\right]^{-2/3}\, .
	\end{aligned}
\end{equation}
One then has to assess the probability that the condition~\eqref{eq:deltac:bc} is fulfilled in a given patch of size $R$ in the universe, in order to compute the abundance of PBHs at every scale. This problem corresponds to the calculation of the mass fraction of PBHs, which we describe in the next section.
\section{Computation of the mass fraction}
\label{sec:MassFraction}
The calculation of the mass fraction of gravitationally-formed objects has received much attention over the last decades, and in this section we describe the main tools that have been developed to address it. While our goal is primarily to identify those that are best suited to the problem at hand in this work, we hope to also clarify the main assumptions that these approaches rest on, and how they are connected together. This section is therefore rather independent from the rest of this article (and is not specific to metric preheating), and while it may serve as a pedagogical introduction to calculations of the mass fraction, it can also be skipped by readers already familiar with these techniques.

In general, cosmological fluctuations can be characterised in terms of an over-density field $\delta(\bm{x})$. In practice, the precise realisation of this field is not known, and one can only predict its statistical properties (for instance, if the field is Gaussian, its is fully characterised by its power spectrum). The problem can thus be divided into two parts. Given a certain realisation of the density field in real space, a first question is to identify the regions where the field will collapse into a certain type of astrophysical object. Second, given the statistical properties of the field, one has to work out the probability that such objects form, and the mass distribution associated to them.

The first question is a very delicate one, and little can be learnt about it analytically without resorting to some approximations. The problem can also be tackled numerically, see \Sec{sec:CompactionFunction} below. Most of the time, it is assumed that a region where the field collapses is one in which the mean density is larger than some threshold value $\delta_\uc$, that may depend on the size of that region (as well as its shape, the details of its surrounding, \etc). In practice, one coarse-grains the field $\delta$ over a spherical region of radius $R$ about the point $\bm{x}$,
\begin{equation}
	\begin{aligned}
\label{eq:def:CoarseGrain}
\delta_{R} (\bm{x}) \equiv \left(\frac{a}{R}\right)^3 \int\dd\bm{y} \delta(\bm{y}) W\left(\frac{a\left\vert \bm{y} - \bm{x}\right\vert}{R} \right),
	\end{aligned}
\end{equation}
and the collapse criterion is often assumed to be of the form $\delta_R>\delta_\uc(R)$. In the above expression, $W$ is a window function such that $W(x)\simeq 1$ if $x\ll 1$ and $W(x)\simeq 0 $ if $x\gg 1$, and normalised in the sense that $ 4\pi \int_0^\infty x^2 W(x)\dd x=1$, such that after coarse graining, a constant field remains a constant field of the same value (here $\bm{x}$ and $\bm{y}$ are comoving spatial coordinates, while $R$ denotes a physical distance, notations are summarised in table~\ref{tab:my_label} for clarity).

The second question is then a well-posed one, and consists in computing the probability that the collapse criterion is satisfied, and the distribution in sizes $R$ (or in resulting mass $M$) of the regions where this is the case, knowing the statistics of the random field $\delta$.
\subsection{The Press-Schechter formalism}
\label{sec:PressSchechter}
A first approach was developed in 1974 by William H. Press and Paul Schechter in \Refa{1974ApJ...187..425P} and proceeds as follows. From \Eq{eq:def:CoarseGrain}, the Fourier transform of the coarse-grained density perturbation is given by
\begin{equation}
	\begin{aligned}
\label{eq:def:tilde:W}
\delta_{R} (\bm{k}) =
\delta(\bm{k})
\underbrace{
4\pi
\left(\frac{a}{kR}\right)^3 \int_0^\infty W\left(\frac{a}{kR} u \right)  \sin(u) u\dd u}_{ \widetilde{W}\left(\frac{k R}{a}\right)}\, ,
	\end{aligned}
\end{equation}
which defines $\widetilde{W}$, that shares similar properties to $W$. Indeed, when $a/(kR) \gg 1$, the values of $u$ such that $W\left(\frac{a}{k R} u \right)$ is not close to zero are much smaller than one, so one can replace $\sin(u)\simeq u$ in the integral over $u$, and using the normalisation condition stated above, one obtains $\widetilde{W}\left[k R/a\right] \simeq 1$ in that limit. In the opposite limit, when $a/(k R) \ll 1$, since $W\simeq 1$ until $u\sim k R/a$, the integral over $u$ in \Eq{eq:def:tilde:W} is of order $k R/a$, hence $\widetilde{W}\left[k R/a\right]\propto [a/(kR)]^2\ll 1 $.\footnote{The details of $\widetilde{W}$ between these two limits depend on those of $W$. For instance, if $W$ is a Heaviside step function,
\begin{equation}
	\begin{aligned}
W(x) = \frac{3}{4\pi} \theta(1-x),
	\end{aligned}
\end{equation}
where $\theta(x)=1$ if $x>0$ and $0$ otherwise, and where the pre-factor is set in such a way that the normalisation condition is satisfied, \Eq{eq:def:tilde:W} gives rise to
\begin{equation}
	\begin{aligned}
\label{eq:Fourier:transform:Heaviside}
\widetilde{W}\left(\frac{k R}{a}\right) = 3 \left(\frac{a}{k R}\right)^3 \left[ \sin \left(\frac{k R}{ H}\right)-\frac{k R}{ H}\cos\left(\frac{k R}{a }\right)\right],
	\end{aligned}
\end{equation}
which verifies the two limits given in the main text.
Conversely, one can set $W$ such that  $\widetilde{W}$ is a Heaviside step function [\ie $W$ is of a similar form as \Eq{eq:Fourier:transform:Heaviside}], and one then has $\delta_{R}(\bm{x}) = \left(2\pi\right)^{-3/2}\int_{k<a/R}\dd {\bm{k}}\delta_{\bm{k}} e^{i\bm{k}\cdot\bm{x}}$.\label{footnote:FilterFunction}}

The power spectrum of $\delta$, $P_\delta$, is defined as
\begin{equation}
	\begin{aligned}
\label{eq:PowerSpectrum:def}
\left\langle \delta(\bm{k}) \delta^*(\bm{k}') \right\rangle = P_\delta(k) \Dirac(\bm{k}-\bm{k}') = \frac{2\pi^2}{k^3} \calP_\delta(k) \Dirac(\bm{k}-\bm{k}')\, ,
	\end{aligned}
\end{equation}
where $\Dirac$ is the Dirac distribution, and which also defines the reduced power spectrum $\calP_\delta$. The power spectrum of the coarse-grained density field, $P_{\delta_R}(k)$, is defined through a similar relation, and \Eq{eq:def:tilde:W} implies that $P_{\delta_R}(k) = P_\delta(k)\widetilde{W}^2(kR/a)$. This allows one to express the coincident two-point function of the coarse-grained density field as
\begin{equation}
	\begin{aligned}
\label{eq:sigmaR}
\sigma^2_R\equiv
\left\langle \delta^2_{R}(\bm{x}) \right\rangle =\int_0^\infty \calP_\delta(k)  \widetilde{W}^2\left(\frac{kR}{a}\right) \frac{\dd k}{k}\, .
	\end{aligned}
\end{equation}
If $\delta$ has Gaussian statistics, so does $\delta_{R}$ since the two are linearly related via \Eq{eq:def:tilde:W}, hence the probability density function associated to $\delta_R$ reads
\begin{equation}
	\begin{aligned}
\label{eq:PDF:deltaR}
P\left(\delta_R\right) = \frac{\ee^{-\frac{\delta^2_R}{2 \sigma_R^2}}}{\sqrt{2\pi \sigma^2_R}}\, .
	\end{aligned}
\end{equation}
This allows one to express the probability that a given region of size $R$ lies above the threshold,
\begin{equation}
	\begin{aligned}
\label{eq:proba:PS:excess}
P\left[\delta_R>\delta_\uc(R)\right] = \int_{\delta_\uc(R)}^\infty P(\delta_R) \dd \delta_R = \frac{1}{2} \mathrm{erfc}\left[\frac{\delta_\uc(R)}{\sqrt{2} \sigma_R}\right]\, .
	\end{aligned}
\end{equation}
\begin{table}[]
    \centering
    \begin{tabular}{l|l}
        Notation & Definition\\
        \hline
        $\delta({\bm x})$ & Density contrast field on a space slice \\
        $\delta_{\bm k}$ & Fourier transform of the density contrast \\
        \multirow{2}*{$\delta_R({\bm x})$} & Density contrast averaged over a patch of (physical) size $R$,\\ & around (comoving) $\bm x$ \\
        \multirow{2}*{$\sigma_R^2=S=\left\langle \delta^2_{R}(\bm{x}) \right\rangle$} & Variance of $\delta_R({\bm x})$ seen as a stochastic variable, \\ & also used to label $R$\\
        $\delta_\uc(R)$ [or $\delta_\uc(S)$] & Collapse criterion over a patch of size $R$ (or with variance $S$) \\
        $\Dirac$ & Dirac delta distribution \\
        $P_\delta(k)$ & Power spectrum of the density contrast\\
        $\calP_\delta(k) = k^3 P_\delta(k) / (2 \pi^2) $ & Reduced power spectrum of the density contrast\\
    \end{tabular}
    \caption{Definitions and notations}
    \label{tab:my_label}
\end{table}
An important remark is that when a given region of size $R$ has an average density above the threshold, it ends up inside a structure, the size of which has to be equal \emph{or larger} than $R$ (for instance if $\delta_R$ is much larger than $\delta_\uc(R)$, by averaging over a slightly larger distance $R'>R$, one may still find $\delta_{R'}>\delta_\uc(R')$, which indicates that the size of the resulting structure is at least $R'$). Therefore, the above probability is the one to lie inside structures of size \emph{at least} $R$.

This naturally leads us to the notion of mass fraction $\beta(M)$, defined such as $\beta(M) \dd \ln M$ corresponds to the fraction of the universe that is comprised in structures of masses between $M$ and $M+\dd M$. By construction, $\int_M^\infty \beta(\tilde{M})\dd \ln \tilde{M}$ corresponds to the fraction of the universe made of structures of sizes larger than $M$. Since there is a one-to-one correspondence between $M$ and $R$ ($M=4\pi \bar{\rho} R^3/3$ at leading order in perturbations), this is nothing but the probability computed in \Eq{eq:proba:PS:excess}. By differentiating both expressions with respect to $M$, one obtains
\begin{equation}
	\begin{aligned}
\label{eq:beta:PS}
\beta(M)=-M\frac{\partial R}{\partial M} \frac{\partial }{\partial R}P\left[\delta_R>\delta_\uc(R)\right].
	\end{aligned}
\end{equation}
Making use of \Eq{eq:proba:PS:excess}, this gives rise to
\bea
\label{eq:beta:PS:explicit}
\beta(M)\dd\ln M = -  
\left(\frac{1}{2}\frac{\delta_\uc}{\sigma_R^2}-\frac{\partial \delta_\uc}{\partial\sigma_R^2}\right) \frac{\ee^{-\frac{\delta_\uc^2}{2\sigma_R^2}}}{\sqrt{2\pi \sigma_R^2}}
\dd \sigma_R^2\, ,
\eea
where we give the result in terms of $\sigma_R^2$ for future convenience. In particular, one finds that the abundance of objects is exponentially suppressed when $\sigma_R$ is smaller than $\delta_\uc(R)$.

Although rather straightforward, this approach is however plagued with the following issue. Consider a region of size $R$ centred on a given point $\bm{x}$, such that the criterion $\delta_R>\delta_\uc(R)$ is \emph{not} satisfied. According to the above considerations, that region is not part of any structure, since it is excluded from \Eq{eq:proba:PS:excess}. However, it could happen that if one considers another radius $R'>R$, the criterion $\delta_{R'}>\delta_\uc(R')$ is satisfied, hence the region of size $R$ is comprised inside a larger region that does collapse into a structure, which contradicts the fact that it is not part of any. This issue is often referred to as the ``cloud-in-cloud problem'', and leads to underestimating the number of structures. It can also be seen by considering the limit $\delta_\uc(R) \to 0$, in which the entire universe should end up in structures. However, letting $\delta_\uc = 0$ in \Eq{eq:proba:PS:excess} leads to only half of the universe lying in collapsed structures. For this reason, the Press-Schechter result is often simply multiplied by $2$. In the following, we will see how to go beyond this approach, and in which cases the Press-Schechter result (with or without the factor $2$) provides a good approximation.
\subsection{The excursion-set approach}
\label{sec:ExcursionSet}
In 1990, Peacock and Heavens proposed in \Refa{Peacock:1990zz} to solve the clould-in-cloud problem of the Press-Schechter formalism using an excursion set approach. They were soon followed in 1991 by Bower, see \Refa{Bower:1991kf}, and by Bond, Cole and  Efstathiou, see \Refa{Bond:1990iw}.

The idea is to view $\delta_{R}$ as a random variable.  When $R$ is very large, recalling that $\widetilde{W}(x)\simeq 0$ when $x\gg 1$, only a small number of modes contribute to \Eq{eq:sigmaR} (namely those for which $k<a/R$), hence $\sigma_R$ is small. In the limit $R\to\infty$, the distribution function~\eqref{eq:PDF:deltaR} thus asymptotes a Dirac distribution centred around zero. Starting from $\delta_{R}=0$ at $R=\infty$, one can then make $R$ decrease. To be explicit, let us consider the case where $\widetilde{W}$ is a Heaviside function,\footnote{In the case where $\tilde{W}$ is taken as a smooth function of the wavenumber, the random noise appearing in the Langevin equation~\eqref{eq:Langevin:sigma2} becomes coloured, which makes the analysis more involved~\cite{Musso:2013pha, Nikakhtar:2018qqg}.} and
\begin{equation}
	\begin{aligned}
\delta_{R}(\bm{x}) = \left(2\pi\right)^{-3/2}\int_{k<\frac{a}{R}}\dd {\bm{k}}\delta_{\bm{k}} e^{i\bm{k}\cdot\bm{x}}\, ,
	\end{aligned}
\end{equation}
see footnote~\ref{footnote:FilterFunction}. As $R$ decreases, more and more modes contribute to the above integral. Each of these modes takes a random realisation, so $\delta_{R}$, seen as a function of $R$, follows a stochastic, Langevin equation, which can be obtained as follows. Between the ``times'' $R$ and $R-\Delta R$, the variation in $\delta_R$ is given by
\begin{equation}
	\begin{aligned}
\delta_{R-\Delta R}(\bm{x}) - \delta_R(\bm{x}) = \left(2\pi\right)^{-3/2}\int_{\frac{a}{R}<k<\frac{a}{R-\Delta R}}\dd {\bm{k}}\delta_{\bm{k}} e^{i\bm{k}\cdot\bm{x}}\, .
    \label{eq:EqualTimeEvaluation}
	\end{aligned}
\end{equation}
Given that $\langle \delta(\bm{k}) \rangle$ vanishes, and since the two-point function of $\delta(\bm{k})$ is given by \Eq{eq:PowerSpectrum:def}, one finds that $\langle \delta_{R-\Delta R}(\bm{x}) - \delta_R(\bm{x}) \rangle = 0 $ and that $\langle [\delta_{R-\Delta R}(\bm{x}) - \delta_R(\bm{x})]^2 \rangle = \calP_\delta(a/R) \Delta R/R $, at leading order in $\Delta R$. This leads to the Langevin equation
\begin{equation}
	\begin{aligned}
\frac{\dd \delta_R(\bm{x})}{\dd R} = \sqrt{\frac{\calP_\delta(a/R)}{R}}\xi(R)\, ,
	\end{aligned}
\end{equation}
where $\xi$ is a white Gaussian noise with vanishing mean and unit variance, \ie $\langle \xi(R) \rangle = 0$ and $\langle \xi(R) \xi(R') \rangle = \Dirac(R-R')$, and one should stress that $R$ is a decreasing variable. Since \Eq{eq:sigmaR} relates $R$ and $\sigma_R^2$ in a monotonous way, the Langevin equation is sometimes written with $S\equiv \sigma_R^2$ as the ``time'' variable, leading to the particularly simple form
\begin{equation}
	\begin{aligned}
\label{eq:Langevin:sigma2}
\frac{\dd \delta_R(\bm{x})}{\dd S}  =  \xi(S)\, ,
	\end{aligned}
\end{equation}
where $\xi(S)$ is a white Gaussian noise normalised with respect to $S$, and where $S$ is an increasing variable.

Starting from $S=0$ (or equivalently, $R=\infty$), the first ``time'' (\ie the largest radius $R$) when $\delta_{R}$ crosses the collapse threshold $\delta_\uc$ corresponds to the size of the largest structure surrounding $\bm{x}$. The calculation thus boils down to solving a first-passage-time problem, for which there are various dedicated techniques in stochastic analysis. Before detailing one of them in \Sec{sec:Volterra}, let us note that if, along a given realisation of the Langevin process~\eqref{eq:Langevin:sigma2}, $\delta_{R}$ crosses $\delta_\uc(R)$ multiple times, then there are as many substructures, but by considering the first crossing time, \ie the largest structure, one accounts for the ``cloud-in-cloud'' mechanism described in \Sec{sec:PressSchechter}.\footnote{The distribution of substructures can also be worked out by solving the ``two-barrier problem''~\cite{Lacey:1993iv}, \ie by deriving the probability that, after upcrossing the threshold $\delta_\uc(R_1)$ at $R=R_1$, a second upcrossing of $\delta_\uc(R_2)$ occurs at $R_2$.}
The distribution of first crossing times, denoted $\Pfpt(S)$ hereafter, thus gives the size (hence mass) distribution of structures, according to\footnote{By comparing \Eqs{eq:beta:PS:explicit} and~\eqref{eq:beta:ExcursionSet}, one can see that the distribution of first crossing times that would be associated to the Press-Schechter result is given by
\bea
\label{eq:Pfpt:PS}
\Pfpt^\mathrm{PS}(S) = \left[\frac{1}{2}\frac{\delta_\uc(S)}{S}-\delta_\uc^\prime(S)\right]\frac{\ee^{-\frac{\delta_\uc^2(S)}{2S}}}{\sqrt{2\pi S}}\, .
\eea}
\begin{equation}
	\begin{aligned}
\label{eq:beta:ExcursionSet}
\beta(M) \dd \ln M = - \Pfpt(S) \dd S \, .
	\end{aligned}
\end{equation}
In this expression, the relationship between $M=4\pi R^3/3$ and $S=\sigma_R^2$ is given by the link between $R$ and $\sigma_R^2$, that is to say by \Eq{eq:sigmaR}, which depends on the initial statistics of the density field.
 \subsection{Volterra integral equations}
 \label{sec:Volterra}
 The first-passage-time problem associated to \Eq{eq:Langevin:sigma2} can be solved by means of a Volterra integral equation, that we derive in this section. We first note that in the absence of any boundary condition, the solution to (the Fokker-Planck equation associated to) \Eq{eq:Langevin:sigma2} is of the Gaussian form
 \begin{equation}
	\begin{aligned}
 \label{eq:Gaussian:PDF:free}
 P_\mathrm{free}(\delta_R,S;\delta_{R,\mathrm{in}},S_\uin) = \frac{1}{\sqrt{2\pi\left(S-S_\uin\right)}}\exp\left[-\frac{\left(\delta_R - \delta_{R,\mathrm{in}}\right)^2}{2\left(S-S_\uin\right)}\right],
	 \end{aligned}
\end{equation}
 which denotes the probability density that the coarse-grained density contrast takes value $\delta_R$ at time $S$, given that at initial ``time'' $S_\uin$, its value is $\delta_{R,\mathrm{in}}$. Since it depends only on $S-S_\uin$ and $\delta_R - \delta_{R,\mathrm{in}}$, hereafter it will be noted as $P(\delta_R - \delta_{R,\mathrm{in}},S-S_\uin)$ for notation convenience. We also introduce $ P(\delta_R,S)$, the solution to \Eq{eq:Langevin:sigma2} when an absorbing boundary at $\delta_R=\delta_\uc(S)$ is enforced, starting from $\delta_R=0$ at $S=0$. It represents realisations of the Langevin equation~\eqref{eq:Langevin:sigma2} that, at time $S$, have not yet crossed out the absorbing boundary. Finally, $\Pfpt(S)$ denotes the probability density associated to the time of first crossing of the boundary $\delta_\uc(R)$, starting from $\delta_R=0$ at $S=0$.

 At a given ``time'' $S$, any realisation of the Langevin equation has either crossed out the absorbing boundary at a previous time $s<S$, or still contributes to the distribution $P$, so one can write
 \begin{equation}
	\begin{aligned}
  \label{eq:Volterra:interm:1}
 1=\int_0^S \Pfpt(s) \dd s + \int_{-\infty}^{\delta_\uc(S)} P(\delta_R,S)  \dd \delta_R\, .
	 \end{aligned}
\end{equation}
 The link between $P$ and $P_\mathrm{free}$ can be derived by noting that, at time $S$, $P$ contains all realisations of $ P_\mathrm{free}$ that have not yet crossed the boundary. In order to get $P(\delta_R)$, one should therefore subtract from $ P_\mathrm{free}$ the probability that a given realisation has crossed the boundary at a previous time, and then, from there, has moved to $\delta_R$. In other words,
 \begin{equation}
	\begin{aligned}
 \label{eq:Volterra:interm:2}
  P(\delta_R,S) =    P_\mathrm{free}(\delta_R,S) - \int_0^{S} \Pfpt(s)P_\mathrm{free}\left[\delta_R-\delta_\uc(s),S-s\right]\dd s\, .
	 \end{aligned}
\end{equation}
 Our goal is to extract $\Pfpt$ from the above two equations. This can be done by differentiating \Eq{eq:Volterra:interm:1} with respect to $S$, and by using \Eq{eq:Volterra:interm:2} to express $P$ in terms of $P_\mathrm{free}$ and $\Pfpt$ only, leading to
 \begin{equation}
	\begin{aligned}
 \Pfpt(S) = & -\delta_\uc'(S) P_\mathrm{free}\left[\delta_\uc(S),S\right]
 +\delta_\uc'(S) \int_0^S\dd s \Pfpt(s) P_\mathrm{free}\left[\delta_\uc(S)-\delta_\uc(s),S-s\right]
 \\ &
 - \int_{-\infty}^{\delta_\uc(S)}\dd\delta_R \frac{\partial}{\partial S} P_\mathrm{free}\left(\delta_R,S\right)
 +\int_{-\infty}^{\delta_\uc(S)}\dd \delta_R \Pfpt(S) P_\mathrm{free}\left[\delta_R-\delta_\uc(S),0\right]
  \\ &
  +\int_{-\infty}^{\delta_\uc(S)}\dd \delta_R \int_0^S\dd s \Pfpt(s) \frac{\partial}{\partial S} P_\mathrm{free} \left[\delta_R-\delta_\uc(s),S-s\right] .
	 \end{aligned}
\end{equation}
 This expression contains 5 terms. The third term can be computed explicitly by making use of \Eq{eq:Gaussian:PDF:free}, and so can the fifth term (where only the integral over $s$ remains). The fourth terms features $P_\mathrm{free}[\delta_R-\delta_\uc,0]$, which is nothing but $\Dirac[\delta_R-\delta_\uc,0]$, and the integral over $\delta_R$ can also be easily performed. This gives rise to
 \begin{equation}
	\begin{aligned}
 \label{eq:Volterra:singularity}
  \Pfpt(S) =& \left[\frac{\delta_\uc(S)}{S}-2\delta_\uc'(S)\right] P_\mathrm{free}\left[\delta_\uc(S),S\right]
  \\ &
  +\int_0^S \dd s \left[2\delta_\uc'(S)- \frac{\delta_\uc(S)-\delta_\uc(s)}{S-s}\right]P_\mathrm{free}\left[\delta_\uc(S)-\delta_\uc(s),S-s\right] \Pfpt(s)\, .
	 \end{aligned}
\end{equation}
 Although mathematically correct, the above expression is nonetheless flawed with a singularity that appears in the integrand of the second term, close to the upper bound of the integral, where it approaches $\delta_\uc'(S) \Dirac(0)$. This leads to numerical issues when trying to solve \Eq{eq:Volterra:singularity} iteratively, which can be dealt with by introducing an averaging procedure when $s\to S$, as proposed for instance in \Refa{Zhang:2005ar}. However, \Eq{eq:Volterra:singularity} is only one version of an infinite set of Volterra equations~\cite{Buonocore:1990vol}, and it can be generalised as follows. Let us consider the realisations of the Langevin equation which, at time $S$, lie at the position $\delta_R=\delta_\uc(S)$. At time $S$, those realisations have necessarily already crossed the boundary, so one can write
 \begin{equation}
	\begin{aligned}
 P_\mathrm{free}\left[\delta_\uc(S),S\right] = \int_0^S\dd s  P_\mathrm{free}\left[\delta_\uc(S)-\delta_\uc(s),S-s\right] \Pfpt(s) .
	 \end{aligned}
\end{equation}
 Multiplying both hands of this equation by a generic function $K(S)$, and plugging the result into \Eq{eq:Volterra:singularity}, one obtains
  \begin{equation}
	\begin{aligned}
 \label{eq:Volterra:kernel}
  \Pfpt(S) =& \left[\frac{\delta_\uc(S)}{S}-2\delta_\uc'(S)+K(S)\right] P_\mathrm{free}\left[\delta_\uc(S),S\right]
  \\ &
  +\int_0^S \dd s \left[2\delta_\uc'(S)- \frac{\delta_\uc(S)-\delta_\uc(s)}{S-s}-K(S)\right]P_\mathrm{free}\left[\delta_\uc(S)-\delta_\uc(s),S-s\right] \Pfpt(s)\, .
	 \end{aligned}
\end{equation}
 Let us stress that this relation is valid for any function $K(S)$. In particular, by setting $K(S)=\delta_\uc'(S)$, one gets rid of the above-mentioned singularity, leading to
   \begin{equation}
	\begin{aligned}
 \label{eq:Volterra:regular}
  \Pfpt(S) =& \left[\frac{\delta_\uc(S)}{S}-\delta_\uc'(S)\right] P_\mathrm{free}\left[\delta_\uc(S),S\right]
  \\ &
  +\int_0^S \dd s \left[\delta_\uc'(S)- \frac{\delta_\uc(S)-\delta_\uc(s)}{S-s}\right]P_\mathrm{free}\left[\delta_\uc(S)-\delta_\uc(s),S-s\right] \Pfpt(s)\, .
	 \end{aligned}
\end{equation}
 This allows one to compute $\Pfpt(S)$ iteratively, by discretising the $S$ variable and starting from $\Pfpt=0$ at $S=0$. See appendix \ref{sec:numerical-volterra} for more details. In \Eq{eq:Volterra:regular}, recall that $P_\mathrm{free}$ is given by \Eq{eq:Gaussian:PDF:free}, and the function $\delta_\uc(S)$, as well as the link between $S$ and $R$ (the physical scale at which the density contrast is coarse grained), are given by the physical details of the problem under consideration.
 \subsection{Relation between the Press-Schechter and excursion-set formalisms}
 \label{sec:Relations:PS:ES}
 The excursion-set approach is an extension of the Press-Schechter formalism, that incorporates the cloud-in-cloud mechanism, and allows for multiple crossings of the threshold value. In this section, in order to clarify the link between the two approaches, we discuss two limiting cases where the excursion set yields a result closely related to the one obtained with the Press-Schechter formalism.
 \subsubsection{Scale-invariant threshold}
 \label{sec:ScaleInvariant:threshold}
When the formation threshold, $\delta_\uc(R)$, does not depend on the scale $R$, the excursion-set approach greatly simplifies. Indeed, in \Eq{eq:Volterra:regular}, the kernel of the integral term vanishes in that limit, so the Volterra implicit equation becomes an explicit formula for the first-passage-time distribution, namely
\begin{equation}
	\begin{aligned}
\label{eq:FPT:PS}
\Pfpt(S)=\frac{\delta_\uc}{S} P_\mathrm{free}\left(\delta_\uc,S\right) =  \frac{\delta_\uc}{S} \frac{\ee^{-\frac{\delta_\uc^2}{2 S}}}{\sqrt{2\pi S}}\, ,
	\end{aligned}
\end{equation}
where, in the second equality, \Eq{eq:Gaussian:PDF:free} has been used. One thus recovers \Eq{eq:Pfpt:PS} exactly, with an additional factor $2$. This proves that the Press-Schechter formula, corrected by the factor $2$ (the origin of which is left rather heuristic in the Press-Schechter approach, see the discussion at the end of \Sec{sec:PressSchechter}), becomes exact in the case of a scale-invariant threshold.
 \subsubsection{Very red threshold}
 \label{sec:RedThreshold}
Another limit of interest is the situation in which the threshold quickly decreases as the scale $R$ decreases. In this case, multiple crossing events become unlikely since after the threshold is crossed for the first time, its value swiftly decays away from the realisation of the overdensity. One therefore expects the Press-Schechter formula (without the additional factor $2$) to be recovered in this regime.

More precisely, this limit can be studied by introducing the rescaled stochastic variable $\hat{\delta}_R(S)\equiv \delta_R(S)/\delta_\uc(S)$, which follows a Langevin equation with a drift term, $\dd \hat{\delta}_R/\dd S = -(\delta_\uc'/\delta_\uc) \hat{\delta}_R + \xi/\delta_\uc$, but with a time-independent threshold since by construction $\hat{\delta}_\uc =1$. The ``very-red-threshold'' limit thus corresponds to the regime in which the drift term dominates over the noise term in this rescaled Langevin equation (conversely, the limit investigated in \Sec{sec:ScaleInvariant:threshold} corresponds to when the noise dominates over the drift).

However, given that, over an infinitesimal time increment $\Delta S$, the drift contribution scales as $\Delta S$ while the typical noise contribution scales as $\sqrt{\Delta S}$, the drift term cannot dominate for arbitrarily small time resolutions. In other words, multiple, repeated crossings are inevitable, and one can only require that they happen within a certain finite time interval, below which we do not try to resolve the distribution of first crossing times. Denoting $\epsilon=\Delta\ln R$ the time resolution one requires [since the mass distribution is usually expressed in $\ln(M)$ units, in practice, one imposes a fixed resolution on $\ln(M)$ or equivalently on $\ln(R)$], the very-red limit can thus be mathematically expressed by requiring that 
\bea
    \forall R_1, R_2 \text{ such that } \ln\left(\frac{R_1}{R_2}\right) > \epsilon, \delta_\uc(R_1) - \delta_\uc(R_2) \gg \sqrt{S(R_2) - S(R_1)}.
    \label{eq:RedCriterion}
\eea
Notice that for practical purposes, it may be enough to satisfy this criterion at the scales $R$ where the mass function is substantial (unless one wants to resolve the tails properly).

In order to see that \Eq{eq:RedCriterion} leads to the Press-Schechter formula, the integral over $s\in [0,S]$ appearing in \Eq{eq:Volterra:regular} can be split into an integral over $[0, S - \eta]$ and  an integral over $[S-\eta, S]$, where $\eta$ is the width of the region where the term $P_\mathrm{free}$ does not lead to an exponential suppression of the integral. From \Eq{eq:Gaussian:PDF:free}, it is order $(\delta_\uc^\prime)^{-2}$, hence it is small in the very-red-threshold limit. It is also related to the $\epsilon$-smoothing scale in $S$-units [so $\eta$ is of order $\epsilon \partial S/\partial\ln R = \epsilon \calP_\delta(k=a/R)$]. In the first integral, the criterion~\eqref{eq:RedCriterion} implies that $P_\mathrm{free}$ exponentially suppresses the integrand, while the second integral is also negligible since the integrand vanishes when $s\to S$ (hence the second integral is of order $\eta^2$). Let us now consider the first term in \Eq{eq:Volterra:regular}. Since $\delta_\uc(S)$ is a decreasing function of $S$ in the regime of interest, the term $P_\mathrm{free}[\delta_\uc(S),S]\sim e^{-\delta_\uc^2(S)/S}$ does not lead to exponential suppression only for large enough values of $S$ such that $S\gtrsim \delta_\uc^2(S)$, which implies that $\delta_\uc/S\lesssim 1/\sqrt{S}$. Since \Eq{eq:RedCriterion} states that $1/\sqrt{\Delta{S}}\ll \vert \delta_\uc^\prime\vert$, this entails that $\delta_\uc/S \ll  \vert \delta_\uc^\prime\vert$, hence \Eq{eq:Volterra:regular} reduces to $\Pfpt(S)\simeq -\delta_\uc^\prime(S) P_\mathrm{free}[\delta_\uc(S),S] $, which matches the Press-Schechter formula~\eqref{eq:Pfpt:PS} in the same limit and without the factor $2$, as announced above. 

Later on in the present work, these considerations will be illustrated by an explicit example in which we will check numerically that when the criterion~\eqref{eq:RedCriterion} is satisfied, the Press-Schechter result is indeed recovered, see \App{sec:NewtonGauge}. 
\subsection{Other methods}
\label{sec:OtherMethods}
Other approaches to the cloud-in-cloud problem have been proposed, and although, in the present work, we make use of the excursion-set method, for completeness, let briefly mention those alternative techniques.
\subsubsection{Supreme statistics}
A first approach to the cloud-in-cloud problem was proposed in 1985 by Bhavsar and Barrow in \Refa{1985MNRAS.213..857B}, and is called the ``supreme statistics'' (or ``extreme-value statistics'') method. The idea is to consider a region of size $R_\mathrm{\ell}$, over which the averaged density contrast is above the threshold. This region is made of $\sim (R_\mathrm{\ell}/R_\mathrm{s})^3$ subregions of size $R_\mathrm{s}$, and one needs to determine the probability that one of these regions is also above the threshold. On generic grounds, considering $n$ samples, each of size $m$, all drawn from the same underlying distribution, the distribution of the maxima within each sample, and therefore the most probable maximum value, can be determined using the supreme statistics (see \Refa{MoradinezhadDizgah:2019wjf} for a recent example of application to PBHs).
\subsubsection{Peak theory}
\label{sec:peak:theory}
In 1986, Bardeen, Bond, Kaiser and Szalay studied the statistics of peaks of Gaussian random fields in \Refa{Bardeen:1985tr}. Assuming that structures form where the density field locally peaks, this allows one to derive the number density of objects satisfying certain conditions on the size of their peak, the volume enclosed within the peak, the deviation from sphericity of the peak, \etc Although the same exponential suppression $\propto \ee^{-\delta_\uc^2(R)/(2 \sigma_R^2)}$ as in the Press-Schechter formalism is obtained, see the discussion below \Eq{eq:beta:PS}, the details of the prefactor are found to be different. \Refa{Bardeen:1985tr} also introduced the ``peak-background split'' approximation, which allows one to compute correlations between peaks belonging to two populations having two different, well-separated scales. The peak-theory and excursion-set approaches can also be combined, see \eg \Refa{Paranjape:2012ks}, and a recent comparison between these different techniques can be found in \Refa{Gow:2020bzo}.
\subsubsection{Improved Press-Schechter formalism}
In 1994, an improved version of the Press-Schechter formalism was proposed by Jedamzik in \Refa{Jedamzik:1994nr}, in which the number density of isolated overdense regions (defined as  overdense regions that are not comprised in larger overdense regions) is computed by making use of Bayes' theorem. The method was improved in  \Refa{Yano:1995gk} (see also \Refa{Nagashima:2001vu} for further refinements, in particular the implementation of the condition that objects form around peaks), and leads to implicit integral equations, similar to the Volterra equations presented in \Sec{sec:Volterra}.
\subsection{Further refinements}
\label{sec:Refinements}
The methods presented above assume that structures form when the overdensity $\delta$ is above a certain threshold $\delta_\uc$, but more refined formation criteria have also been studied.
\subsubsection{Critical scaling}
From studying spherically symmetric collapse of a massless scalar field by numerical means, Choptuik has shown in \Refa{Choptuik:1992jv} that, close to the critical threshold, the mass of the resultant black hole, $\mPBH$, is proportional to $ (\delta-\delta_\uc)^\gamma$ where $\gamma\simeq 0.37$ is a universal exponent. This has been generalised to radiation fluids in \Refs{Evans:1994pj, Koike:1995jm}, and reviews of critical phenomena in gravitational collapse can be found in \Refs{Gundlach:1999cu, Gundlach:2002sx}. The relation
\begin{equation}
	\begin{aligned}
\label{eq:critical:collapse}
\mPBH = K m_H (\delta-\delta_\uc)^\gamma,
	\end{aligned}
\end{equation}
where $K$ is a constant and $m_H$ is the mass contained within a Hubble volume at the time the black holes form,
has been applied to the calculation of the PBH mass fraction in \Refa{Niemeyer:1997mt}. It was then numerically analysed in the context of PBHs in \Refs{Musco:2008hv, Musco:2012au}, and further investigations on its applicability can be found in \Refa{Byrnes:2018clq}.
\subsubsection{Compaction function}
\label{sec:CompactionFunction}
Recent numerical works by Musco, see \Refa{Musco:2018rwt} and subsequent publications, suggest that a more accurate criterion for PBH formation follows from the analysis of the compaction function
\begin{equation}
	\begin{aligned}
C(r)=2 \frac{m(r,t)-\bar{m}(r,t)}{R(r,t)},
	\end{aligned}
\end{equation}
where $r$ is the distance away from the overdensity peak, $m(r,t)-\bar{m}(r,t)$ is the excess mass contained inside a sphere of radius $r$, and $R(r,t)$ is the areal radius. The scale of the fluctuations relevant for the formation of PBHs is the one that maximises the compaction function. In other words, a PBH forms at the scale $r_\mathrm{m}$ where $C(r)$ is maximal, provided $C(r_{\mathrm{m}})$ is larger than some threshold $C_\uc$ (which is roughly equivalent to requiring that the overdensity averaged over a sphere of radius $r_{\mathrm{m}}$ overcomes the threshold value $\delta_\uc$, see \Refa{Young:2019osy}).

Critical collapse can be implemented with the criterion based on the compaction function (see Appendix A of \Refa{Young:2020xmk}), simply by replacing $\delta$ by $C(r_{\mathrm{m}})$, and $\delta_\uc$ by $C_\uc$ in \Eq{eq:critical:collapse}, and by using different values of the constants $K$ and $C_\uc$ ($\gamma$ is still the same).

Compaction-function based criteria have also been employed within the peak-theory approach in \Refa{Young:2020xmk} (see also \Refs{Suyama:2019npc, Germani:2019zez}). They cannot be directly implemented in the excursion set program since, here, the size of the structure is not determined by the first $C$-crossing of the threshold, but rather by the ``time'' (\ie scale) at which $C$ is maximal (and by the value of that maximum through the critical-scaling relation). See also \Refa{Young:2019osy} for a comparison between these different criteria.
\subsection{Application to primordial black holes}
\label{sec:MassFraction:PBHs}
The methods introduced above were originally developed in the context of large-scale structures in general (with the exception of the refinements presented in \Sec{sec:Refinements}), and their application to the calculation of the mass distributions of PBHs requires some further considerations.
\subsubsection{Removing the super-horizon modes}
\label{sec:Remove:Super:Horizon}
Primordial black holes are expected to form when a large curvature fluctuation re-enters the Hubble radius after inflation, and collapses into a black hole. The relevant smoothing scale $R$ is therefore the Hubble radius at the time the black hole forms. In the coarse-graining procedure~\eqref{eq:def:CoarseGrain}, given the properties of the function $\widetilde{W}$ detailed at the beginning of \Sec{sec:PressSchechter}, most modes $k$ that contribute to \Eq{eq:sigmaR} are such that $k<a/R$, hence they are super Hubble at the time of formation. This raises two issues (that do not appear in the context of large-scales structures, where all modes that contribute to the overdensity field are far inside the Hubble radius).

First, far super-Hubble curvature perturbations should only lead to a local rescaling of the background metric, and can hardly determine whether or not an object forms inside a local Hubble patch. Second, in general relativity, there is no unique definition of the energy density (hence of the density contrast), which depends on the space-like hypersurface on which it is computed.  All possible choices coincide on sub-Hubble scales (where observations are performed), but they differ on super-Hubble scales. In practice, in most gauges studied in the literature, the density contrast $\delta$ and the Bardeen potential $\Phi$ are related through a formula of the form~\cite{Malik:2008im}
\begin{equation}
	\begin{aligned}
\label{eq:delta:Phi}
\delta_k =  -\frac{2}{3} \left(\frac{k}{aH}\right)^2\Phi_k + \alpha \Phi_k + \beta \frac{\dot{\Phi}_k}{H}\, ,
	\end{aligned}
\end{equation}
where $\alpha$ and $\beta$ are two constants, that possibly depend on the equation-of-state parameter $w$. For instance, in the Newtonian gauge, $\alpha=\beta=-2$, in the flat gauge, $\alpha=5-3w$ and $\beta=-2$, and in the comoving gauge, $\alpha=\beta=0$. On sub-Hubble scales, when $k\gg aH$, the first term in \Eq{eq:delta:Phi} dominates, and $\delta_k$ does not depend on the choice of slicing as mentioned above. Since the Bardeen potential $\Phi$ is related to the curvature perturbation $\zeta$ via~\cite{Wands:2000dp}
\begin{equation}
	\begin{aligned}
\label{eq:zeta:Bardeen}
\zeta_k  = \frac{2}{3}\frac{\dot{\Phi}_k/H+\Phi_k}{1+w}+\Phi_k\, ,
	\end{aligned}
\end{equation}
on super-Hubble scales, where $\dot{\Phi}_k$ can be neglected since it is proportional to the decaying mode, one has
\begin{equation}
	\begin{aligned}
\label{eq:delta:Phi:MD}
\delta_k \simeq  -\frac{2}{5} \left(\frac{k}{aH}\right)^2\zeta_k + \frac{3\alpha}{5} \zeta_k \, .
	\end{aligned}
\end{equation}
This shows that there are essentially two families of slicings for the density contrast. If $\alpha\neq 0$, as in the Newtonian gauge or in the flat gauge (if $w\neq 5/3$) for instance, on super-Hubble scales, $\delta_k \propto \zeta_k$, so for quasi scale-invariant curvature power spectra (as expected from inflation), super-Hubble modes give a substantial contribution to \Eq{eq:sigmaR}, leading to the problem mentioned above. If, on the contrary, $\alpha=0$, as in the comoving gauge, then $\delta_k \propto k^2 \zeta_k$ on super-Hubble scales, hence it is highly suppressed and far super-Hubble modes do not contribute much to the integral of \Eq{eq:def:tilde:W}. For this reason, in \Refa{Young:2014ana}, it is proposed to work with the comoving density contrast, as a way to effectively remove the contribution from far super-Hubble modes. In principle, a well-defined formation criterion should come with a prescription for which density contrast to use, and in practice, formation criteria derived from numerical investigations are indeed most often formulated in the comoving slicing, see for instance \Refs{Harada:2015yda, Musco:2018rwt}. This is why we will adopt this choice in what follows, and in \App{sec:NewtonGauge}, we will investigate how the results are modified if one makes a different choice and works with the Newtonian density constrast.
\subsubsection{Simplified Press-Schechter estimate}
\label{sec:simplified:PS}
The above considerations lead to a simplified version of the Press-Schechter formalism that is often used to estimate the abundance of PBHs. Indeed, in the integral of \Eq{eq:sigmaR}, modes that lie far below the coarse-graining scale, such that $k\gg a/R$, are cut away by the filter function $\widetilde{W}$, while modes that lie far above the coarse-graining scale, such that $k\ll a/R$, give a negligible contribution since they are suppressed by $k^2$ as explained in \Sec{sec:Remove:Super:Horizon}. Therefore, the only modes that give a substantial contribution to \Eq{eq:sigmaR} are those such that $k\sim a R$, and \Eq{eq:sigmaR} can be approximated as
\begin{equation}
	\begin{aligned}
\sigma_R^2 \sim \calP_\delta\left(k=\frac{a}{R}\right) .
	\end{aligned}
\end{equation}
It is then usually assumed that the power spectrum peaks at a single scale. At the time when that scale re-enters the Hubble radius, the fraction of Hubble patches where the density contrast is larger than the critical value is given by \Eq{eq:proba:PS:excess}, so it is common to use directly use \Eq{eq:proba:PS:excess} as the mass fraction, and to write (see for instance \Refa{Harada:2013epa})
\begin{equation}
	\begin{aligned}
\label{eq:beta:PS:simplified}
\beta(M) \sim \erfc\left[\frac{\delta_\uc (R)}{\sqrt{2 \calP_\delta (k)}}\right] \sim \frac{\sqrt{2 \calP_\delta}}{\sqrt{\pi} \delta_\uc}\ee^{-\frac{\delta_\uc^2}{2 \calP_\delta}},
	\end{aligned}
\end{equation}
where $R$ is the Hubble radius and $M$ the Hubble mass at the time when $k=aH$, and where in the last expression, we have expanded the error function in the regime where PBHs are rarely produced, \ie when $\delta_\uc \gg \sqrt{\calP_\delta}$. Let us note that, when comparing \Eq{eq:beta:PS:simplified} to the full Press-Schechter formula~\eqref{eq:beta:PS}, the same exponential suppression is obtained, but the prefactor is obviously different. However, as noted in \Sec{sec:peak:theory}, that prefactor was also found to differ in other approaches such as peak theory, so one may consider that the details of the prefactor are ultimately dependent on the approach one follows, and that only the exponential suppression is a robust result, in which case \Eq{eq:beta:PS:simplified} may provide a useful estimate. This is why this formula is widely employed, including in \Refa{Martin:2019nuw} for the calculation of PBHs from metric preheating, in which we are interested in this work.

However, it is pretty clear that this approximation breaks down for broad spectra, \ie when a wide range of scales is involved in the formation of PBHs. As pointed out \eg in \Refa{Suyama:2019npc}, the problem is that \Eq{eq:proba:PS:excess} is not a differential quantity. So it can happen for instance that the integrated mass fraction,
\begin{equation}
	\begin{aligned}
\label{eq:Omega:PBH}
\OmegaPBH = \int_0^\infty \beta(M) \dd \ln M\, ,
	\end{aligned}
\end{equation}
which corresponds to the fraction of the total energy budget that is comprised inside PBHs, is found to be larger than one if \Eq{eq:beta:PS:simplified} is used, which clearly signals an inconsistency. This is in fact precisely what happens in \Refa{Martin:2019nuw}, where renormalisation procedures had to be introduced to cope with this issue, see \App{sec:Comparison:Previous:Work}. In the rest of this paper, we re-examine this problem with the excursion-set formalism, in order to assess more precisely the mass distribution of PBHs produced from the preheating instability.
\section{Primordial black holes from metric preheating}
\label{sec:PBH}
We are now in a position in which we can apply the excursion-set formalism, presented in \Secs{sec:ExcursionSet} and~\ref{sec:Volterra}, to the calculation of the mass fraction of PBHs arising from the preheating instability described in \Sec{sec:MetricPreheating}.
\subsection{Collapse criterion}
\label{sec:Collapse:Criterion}
As explained in \Sec{sec:ExcursionSet}, one of the two ingredients required by the excursion set approach is the critical value of the density contrast, above which PBHs form. This was derived in the case of metric preheating in \Sec{sec:MetricPreheating}, see \Eq{eq:deltac:bc}. This formula provides a critical value for the density contrast evaluated at a time that depends on the scale $R$ (namely at the band-crossing time of $R$). Below, we rather choose to express all relevant quantities at the same reference time.
The reason is that if the time of evaluation depends on the scale $R$, then, when deriving \Eq{eq:EqualTimeEvaluation}, an additional term, which stands for the time evolution of the mode function, appears.
This term acts as a drift term in the Langevin equation~\eqref{eq:Langevin:sigma2}, but it can be absorbed by a change of variables.
This amounts to rescaling $\delta$ with a transfer function that evolves the density contrast to a fixed time, hence is equivalent to working with fixed-time quantities anyway. 
In either case, one has to relate the value of the density contrast at the band-crossing time to its value at a fixed, reference time.

Two natural choices for such a reference time are (i) the time at the end of inflation and (ii) the time at the end of the instability phase. The problem with the latter choice is that, since coarse graining $\delta$ at a scale $R$ selects modes that are larger than $R$, see the discussion at the beginning of \Sec{sec:Remove:Super:Horizon}, the scales that contribute to $\delta_R$ at the end of the instability are either sub- or super-Hubble, hence they have \apriori different behaviours (\ie different transfer functions) between the band-crossing time of $R$ and the end of the instability, which makes it harder to relate $\delta_R[t_\mathrm{bc}(R)]$ with $\delta_R(t_\Gamma)$. The first option is therefore more convenient, since all scales larger than $R$ are super-Hubble between the end of inflation and the band-crossing time of $R$.

One thus has to study how $\delta$ evolves on super-Hubble scales during the oscillatory phase. In \Sec{sec:Remove:Super:Horizon}, we explained why the comoving density contrast had to be considered. From \Eq{eq:delta:Phi}, it is related to the Bardeen potential via $\delta_{\bm{k}} = -2/3 [k/(a H)]^2 \Phi_{\bm{k}}$, while the Bardeen potential is related to the curvature perturbation through \Eq{eq:zeta:Bardeen}. As shown in \Sec{sec:MetricPreheating}, the curvature perturbation $\zeta_{\bm{k}}$ is constant on super-Hubble scales during the oscillatory phase. As a consequence, if the equation-of-state parameter $w$ were constant, then \Eq{eq:zeta:Bardeen} seen as a differential equation for $\Phi_{\bm{k}}$ would show that, up to a quickly decaying mode, $\Phi_{\bm{k}}$ reaches a constant value, hence $\delta_{\bm{k}}\propto (a H)^{-2}$.

\begin{figure}[t]
\begin{center}
  \includegraphics[width=1.0\textwidth]{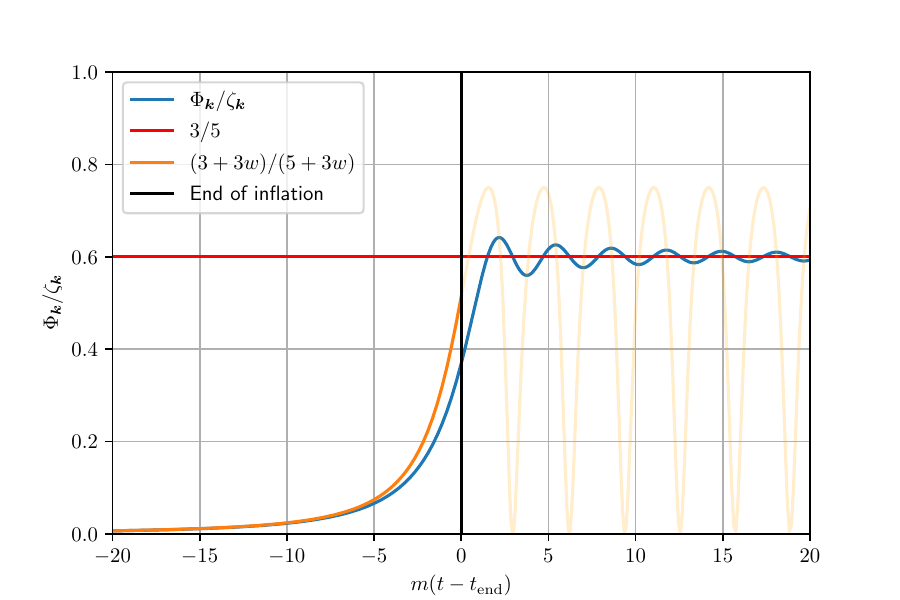}
\caption{Bardeen potential $\Phi_{\bm{k}}$ rescaled by the curvature perturbation $\zeta_{\bm{k}}$ during the last \efolds~of inflation and the first \efolds~of the oscillatory phase, in the same situation as the one displayed in \Fig{fig:scale}, for a scale ${\bm{k}}$ that is sufficiently far outside the Hubble radius such that $\zeta_{\bm{k}}$ can be taken as constant. The blue line stands for the full numerical solution of \Eq{eq:zeta:Bardeen}, seen as a differential equation for $\Phi_{\bm{k}}(t)$, where $w(t)$ and $H(t)$ are extracted from \Fig{fig:scale}. The red line stands for the approximation~\eqref{eq:Phi:zeta:Appr}, $\Phi_{\bm{k}}/\zeta_{\bm{k}} = 3/5$, obtained as the late-time solution of \Eq{eq:zeta:Bardeen} when setting $w=0$ and $H=2/(3 t)$, and towards which the full numerical result asymptotes after a few oscillations. The orange line stands for \Eq{eq:zeta:Bardeen} where we neglected $\dot{\Phi}_{\bm{k}}/H$ with respect to $\Phi_{\bm{k}}$. This approximation is well justified on super-Hubble scales during inflation, since $w$ is almost constant there, but fails during the subsequent oscillatory phase where $w$ vanishes on average but otherwise undergoes large oscillations.}
\label{fig:BardeenOscillations}
\end{center}
\end{figure}
However, during the oscillatory phase, $w$ is only constant on average, and otherwise undergoes large oscillations. Indeed, at leading order in $H/m$, the inflaton oscillates according to $\phi\simeq \phi_\uend (a_\uend/a)^3\sin(m t)$, which gives rise to $w= \cos(2 m t) + \order{H/m}$. Through \Eq{eq:zeta:Bardeen}, these oscillations give rise to oscillations in $\Phi_{\bm{k}}$ with frequency $\sim m$, hence $\Phi'/(H \Phi)$ is of order $m/H\gg 1$ and can a priori not be neglected in \Eq{eq:zeta:Bardeen}. Nonetheless, deep in the oscillatory phase, when $H\ll m$, these oscillations can be averaged out, and one has
\begin{equation}
	\begin{aligned}
\label{eq:Phi:zeta:Appr}
\Phi_{\bm{k}} \simeq \frac{3}{5}\zeta_{\bm{k}}\, .
	\end{aligned}
\end{equation}
In order to verify the validity of this statement, in \Fig{fig:BardeenOscillations}, we display the numerical solution of \Eq{eq:zeta:Bardeen} in the same situation as the one shown in \Fig{fig:scale} (namely, from the numerical solution of the Klein-Gordon equation, we extract $w(t)$ and $H(t)$, and solve \Eq{eq:zeta:Bardeen} for $\Phi_{\bm{k}}(t)$ while assuming that $\zeta_{\bm{k}}$ is constant). We also superimpose the approximation~\eqref{eq:Phi:zeta:Appr}. One can see that, after a few oscillations, it provides an excellent fit to the full numerical solution. Therefore, on average, the Bardeen potential is indeed constant, and $\delta_{\bm{k}}\propto (a H)^{-2}$.

As stressed above, all scales contributing to $\delta_R$ are super-Hubble between the end of inflation and the band-crossing time, hence they all evolve according to $\delta_{\bm{k}}\propto (a H)^{-2}\propto a$ in a matter-dominated era. Therefore, $\delta_R$ itself evolves in the same way, and
\begin{equation}
	\begin{aligned}
\delta_R\left(t_\uend\right) = \delta_R\left[t_{\mathrm{bc}}(R)\right] \frac{a_\uend}{a\left[t_{\mathrm{bc}}(R)\right]}\, .
	\end{aligned}
\end{equation}
Combining this result with \Eq{eq:deltac:bc}, and denoting by $R_\uend$ the value of $R$ at the end of inflation, one obtains for the critical value of the density contrast at the end of inflation
\begin{equation}
	\begin{aligned}
\label{eq:collapse:threshold:end}
\delta_\uc(R,t_\uend) = \left(\frac{3\pi}{2}\right)^{2/3} \left[\frac{H_\uend}{H_\Gamma}-\left(R_\uend H_\uend \right)^{3}\right]^{-2/3}\, .
	\end{aligned}
\end{equation}

\begin{figure}
	\centering
	\includegraphics[width=\textwidth]{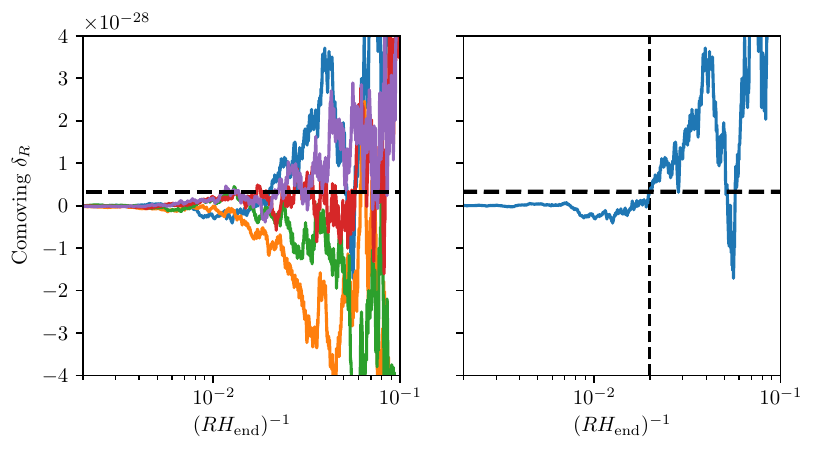}
	\caption{Example of Langevin trajectories for the density contrast evaluated on comoving slices at the end of inflation, and coarse-grained at the scale $R$, for $H_\uend = 10^{-8} \Mp$ and $H_\Gamma = 10^{-25} \Mp$. The (quasi) horizontal black dashed line shows the collapse criterion~\eqref{eq:collapse:threshold:end}. In the right panel, we isolate one realisation and the vertical dashed line denotes the first crossing ``time'' (\ie scale) of the critical threshold.}
	\label{fig:monte_carlo}
\end{figure}%

\subsection{Overdensity variance}
\label{sec:OverDensityVariance}
The second ingredient required by the excursion-set approach is the expected variance of $\delta_R$, $\sigma_R^2$, and how it relates to $R$. If one sets the function $\widetilde{W}$ to a Heaviside function in \Eq{eq:sigmaR}, see footnote~\ref{footnote:FilterFunction}, one has
\begin{equation}
	\begin{aligned}
\sigma_R^2(t_\uend) = \int_0^{a/R}\calP_\delta\left(k,t_\uend\right) \frac{\dd k}{k}\, .
	\end{aligned}
\end{equation}
As explained in \Sec{sec:Collapse:Criterion}, in the comoving gauge, the density contrast is related to the Bardeen potential via $\delta_{\bm{k}} = -2/3 [k/(a H)]^2 \Phi_{\bm{k}}$, and the link between the Bardeen potential and the curvature perturbation is given by \Eq{eq:Phi:zeta:Appr}. This gives rise to
\begin{equation}
	\begin{aligned}
\label{eq:sigma2R:Pzeta}
\sigma_R^2(t_\uend) =  \left(\frac{2}{5}\right)^2\int_0^{a/R} \left(\frac{k}{a_\uend H_\uend}\right)^4   \calP_\zeta\left(k,t_\uend\right) \frac{\dd k}{k}\, .
	\end{aligned}
\end{equation}
In this expression, the curvature power spectrum $\calP_\zeta\left(k,t_\uend\right) $ depends on the details of the inflationary phase that precedes metric preheating. Hereafter, we will assume that the inflationary potential can still be assumed as being almost quadratic at the time the scales of interest cross out the Hubble radius during inflation, where the mass of the inflaton is related to $H_\uend$, and the power spectrum is obtained by solving the Mukhanov-Sasaki equation~\eqref{eq:eomv} numerically during inflation, starting from the Bunch-Davies vacuum.

\begin{figure}
    \centering
    \includegraphics{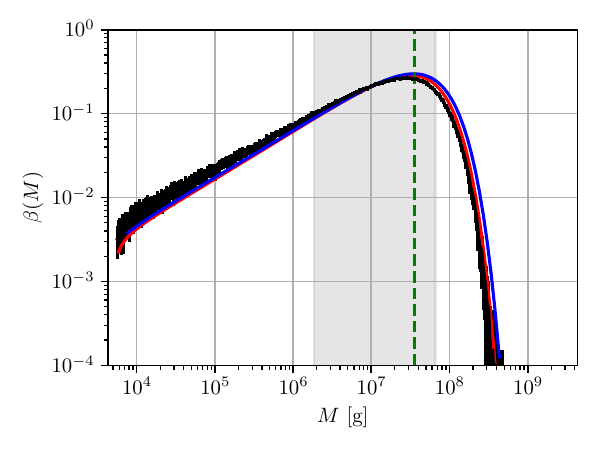}
    \caption{Mass fraction $\beta$ of primordial black holes for $H_\uend = 10^{-8} \Mp$ and $H_\Gamma = 10^{-25} \Mp$, as a function of the mass $M$ in grams. The vertical black bars stand for the distribution of first crossing times obtained from $10^6$ simulated realisations of the Langevin equation~\eqref{eq:Langevin:sigma2}, binned into $1000$ logarithmically spaced values of $R$. The size of the bars correspond to $5\sigma$ estimates of the statistical error by jackknife resampling. The red line corresponds to numerically solving the Volterra equation~\eqref{eq:Volterra:regular}, using the method described in \App{sec:numerical-volterra}. The blue line displays the analytical approximation developed in \Sec{sec:AnalyticalApproximation}, which provides a good fit to the full numerical. The vertical green line denotes the mass at which $\beta$ peaks, as estimated from \Eq{eq:meanMass}, and the grey shaded area stands for the $1\sigma$ deviation of $\ln(M)$ according to the distribution $\beta(M)$, centred on its mean value.}
    \label{fig:mass-function}
\end{figure}
\subsection{Numerical results}
\label{sec:NumericalResults}

As explained in \Sec{sec:ExcursionSet}, in the excursion-set approach, the mass fraction of PBHs is directly related to the first-crossing-time distribution of realisations of the Langevin equation~\eqref{eq:Langevin:sigma2}, see \Eq{eq:beta:ExcursionSet}. This distribution can be estimated using a Monte-Carlo sampling. In \Fig{fig:monte_carlo}, we show a few realisations of the Langevin equation~\eqref{eq:Langevin:sigma2}, for the density contrast $\delta_R$ evaluated on comoving slices at the end of inflation. The collapse threshold obtained in \Eq{eq:collapse:threshold:end} is displayed with the (quasi) horizontal black dashed line, and in the right panel, the time of first crossing is shown with the vertical black dashed line.

This gives rise to the mass fraction displayed in \Fig{fig:mass-function} with the vertical black bars and where the size of the bars corresponds to a $5\sigma$ estimate of the statistical error using jackknife resampling. We report a good convergence in estimating the mass fraction for a sample of $10^6$ trajectories with $1000$ logarithmically spaced values of $R$.

This method is, however, computationally expensive, especially in the tails of the distribution where one needs to simulate a very large number of Langevin realisations to compensate for the sparse statistics. Instead, as explained in \Sec{sec:Volterra}, one can solve the Volterra equation~\eqref{eq:Volterra:regular}, making use of the numerical procedure outlined in \App{sec:numerical-volterra}.
For $\np$ values of $R$, this algorithm requires to invert a $\np \times \np$ lower triangular matrix, which is far more efficient than having to solve Langevin realisations.\footnote{In terms of numerical performance, when producing \Fig{fig:mass-function} we found on our machine that the Volterra approach is more than $1000$ times faster than the Monte-Carlo sampling of Langevin realisations. This is a generic result: if $N_{\mathrm{real}}$ realisations are simulated, the average number of points per bin is of order $\bar{n}\sim N_{\mathrm{real}}/\np$, hence the statistical error is of order $1/\sqrt{\bar{n}}\sim \sqrt{\np/N_\mathrm{real}}$. Requiring that this is smaller than a target accuracy $\epsilon$ leads to $N_\mathrm{real}>\np/\epsilon^2$. Since each realisation requires $\np$ evaluations of the noise, the Langevin approach relies on $\sim \np^2/\epsilon^2$ numerical operations. On the other hand, the Volterra method implies to invert a $\np\times\np$ triangular matrix (see \App{sec:numerical-volterra}), which requires $\sim \np^2$ operations (using the ``forward substitution'' algorithm), and is thus more efficient by a factor $\epsilon^2$.} In practice, we find good convergence for $\np \geq 500$. The result is displayed in \Fig{fig:mass-function} with the red line, where one can check that the two methods give compatible results.

\begin{figure}[t]
\begin{center}
  \includegraphics[width=0.445\textwidth,clip=true]{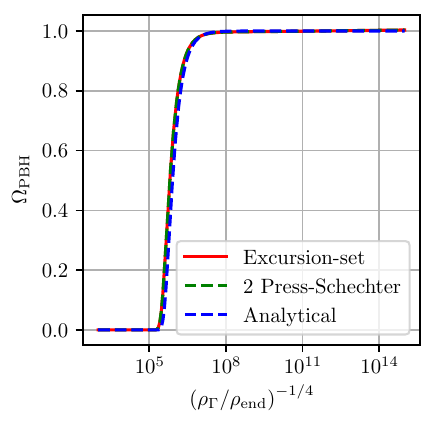}
  \includegraphics[width=0.545\textwidth,clip=true]{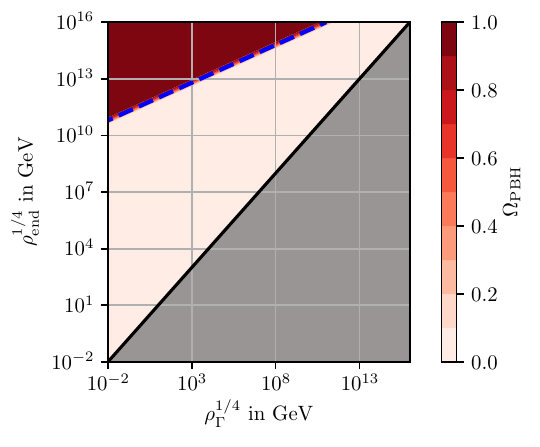}
\caption{Total fraction of the universe comprised in PBHs, $\OmegaPBH$, as a function of $\rho_\mathrm{end}$, the energy density at the end of inflation, and $\rho_\Gamma$, the energy density at the end of the instability phase. On the left panel, we fix $\rho_\uend=10^{-12}\Mp^4$ and let $\rho_\Gamma$ vary. The solid red curve is the full numerical result obtained in the excursion-set approach. The dashed green line corresponds to the Press-Schechter result with the additional factor $2$, which becomes exact in the limit of a scale-invariant threshold, see \Sec{sec:ScaleInvariant:threshold}. The dashed blue line corresponds to the analytical approximation~\eqref{eq:Omegatot:anal}. On the right panel, the full parameter space is explored (where $\rho_\Gamma<\rho_\uend$ since the oscillatory phase occurs after inflation). The colour encodes the value of $\OmegaPBH$, and the transition from tiny values to values close to one is very abrupt. The dashed blue line stands for the analytical estimate~\eqref{eq:bound:Omegatot_eq_1} for the location of this transition.}
\label{fig:comoving}
\end{center}
\end{figure}

The total fraction of the universe made of PBHs, $\OmegaPBH$, is obtained by integrating the mass fraction, see \Eq{eq:Omega:PBH}. The result is shown in \Figs{fig:comoving} as a function of $\rho_\mathrm{end}$, the energy density at the end of inflation, and $\rho_\Gamma$, the energy density at the end of the instability phase. It is obtained from numerically solving the Volterra equation. One can see that the transition from small values of $\OmegaPBH$ to values of order one is very sharp, and that there exist a region in parameter space, corresponding to the dark red region in the right panel of \Fig{fig:comoving}, where the universe is dominated by a gas of PBHs already at the end of the oscillatory phase. 

\begin{figure}
    \begin{subfigure}{0.49\textwidth}
    	\includegraphics[width=\textwidth]{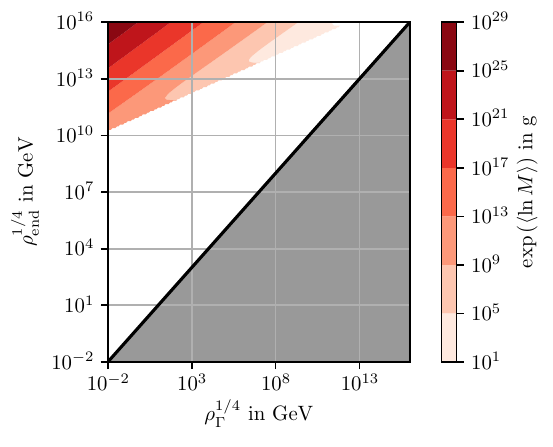}
    	\caption{Mean PBH mass}
    	\label{fig:comoving-mean}
    \end{subfigure}
    \begin{subfigure}{0.49\textwidth}
    	\includegraphics[width=\textwidth]{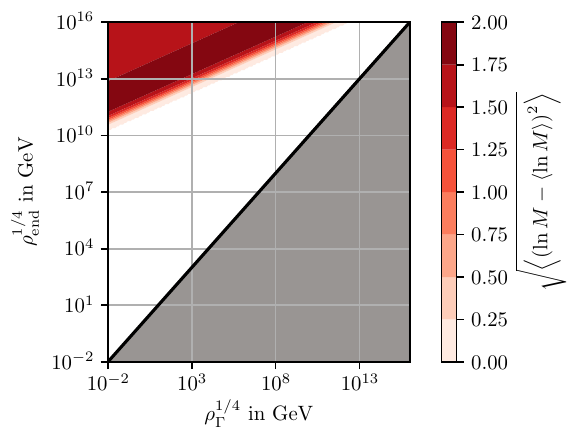}
    	\caption{Dispersion of the PBH masses}
    	\label{fig:comoving-dispersion}
    \end{subfigure}
    \caption{Typical masses of the PBHs produced in metric preheating. The left panel shows the average mass, computed from the mass fraction distribution. In the white region the abundance of PBHs is too small to be numerically resolved, hence the average mass cannot be computed. On the right panel, we show the standard deviation of $\ln(M)$, which describes the typical width of the mass fraction distribution.}
    \label{fig:masses}
\end{figure}

Another quantity of interest is the typical mass of the resulting black holes, which is displayed across parameter space in \Fig{fig:masses}. In the left panel, the average mass is shown, in the form of $\exp(\langle \ln M \rangle)$, and one can see that in the region of parameter space in which PBHs are substantially produced, it spans  a large range of values, from $10$g to $10^{33}\mathrm{g}\sim M_\odot$, where $M_\odot$ is the mass of the sun. In the right panel, the standard deviation of $\ln (M)$ is displayed, in order to see how many orders of magnitude the mass fraction distribution covers (this standard deviation is also shown with the grey shaded stripe in \Fig{fig:mass-function}). One can see that $\Delta\ln M<2$, so the mass distributions never extend over many orders of magnitude.

\subsection{Analytical approximation}
\label{sec:AnalyticalApproximation}
In this section, we try to gain further analytical insight into the numerical results presented in \Sec{sec:NumericalResults} by performing a few approximations. The main simplification comes from the remark that in \Eq{eq:collapse:threshold:end}, $H_\uend/H_\Gamma\gg (R_\uend H_\uend)^3$ except for the scales that enter the instability band close to the end of the oscillatory phase. Those undergo little amplification anyway, and are therefore mostly irrelevant for PBH production. In that limit, one can approximate \Eq{eq:collapse:threshold:end} as
\begin{equation}
	\begin{aligned}
	\label{eq:deltac:constant}
\delta_\uc(R,t_\uend)\simeq \left(\frac{3\pi}{2}\frac{H_\Gamma}{H_\uend}\right)^{2/3}\, ,
	\end{aligned}
\end{equation}
which crucially does not depend on $R$ (hence on $S$) anymore. As explained in \Sec{sec:ScaleInvariant:threshold}, in that limit, the first passage-time distribution is given by \Eq{eq:FPT:PS} and one recovers the result from the Press-Schechter formalism, with an additional factor $2$. Note that this simplification is rather coincidental in the present case, since when expressed at the band-crossing time, the threshold is strongly scale dependent, but the behaviour of the comoving density contrast on super-Hubble scale is such that, when re-expressed at the end of inflation, it exactly cancels out that scale dependence (this is no longer true if other density contrasts are used, see \App{sec:NewtonGauge}).

We have not displayed the (two times) Press-Schechter formula~\eqref{eq:FPT:PS} in \Fig{fig:mass-function} since it cannot be distinguished by eye from the red line (\ie the full numerical result), confirming that this is indeed an excellent approximation. The (two times) Press-Schechter formula is, however, shown in the left panel of \Fig{fig:comoving} as the dashed green line, where one can check that it reproduces the excursion set result very accurately.

Then, in order to make \Eq{eq:FPT:PS} explicit, one needs to relate the overdensity dispersion $S=\sigma_R^2$ to the mass $M$, which implies to first derive an approximation for \Eq{eq:sigma2R:Pzeta}. In single-field inflation, in the slow-roll approximation, one has~\cite{Mukhanov:1985rz, Mukhanov:1988jd}
\begin{equation}
	\begin{aligned}
\label{eq:Pzeta:single_field:slow_roll}
	\calP_\zeta(k) \simeq \frac{H^2(k)}{8\pi^2\Mp^2\epsilon_1(k)}\, ,
	\end{aligned}
\end{equation}
where $H(k)$ and $\epsilon_1(k)$ are respectively the values of the Hubble parameter and the first slow-roll parameter when the scale $k$ crosses-out the Hubble radius during inflation. As argued before, towards the end of inflation when the inflaton approaches the minimum of its potential, the potential can be approximated as being quadratic, $V(\phi)\simeq m^2\phi^2/2$. In such a potential, the slow-roll trajectory reads~\cite{Martin:2013tda} $\phi(k) \simeq \Mp\sqrt{2-4(N-N_\uend)} $. Given that, still at leading order in slow roll, one has $H^2\simeq V/(3\Mp^2)$ and $\epsilon_1\simeq (V'/V)^2/(2\Mp^2)$, this allows one to approximate the power spectrum~\eqref{eq:Pzeta:single_field:slow_roll} as $\calP_\zeta \simeq 3H_\uend^2/(24 \pi^2 \Mp^2)[1-2\ln (k/a_\uend H_\uend)]^2$. Since we have made use of the slow-roll approximation, which breaks down when inflation ends, this formula is in fact accurate only for scales that emerge sufficiently early before the end of inflation, $k\ll a_\uend H_\uend$, and in this regime, \Eq{eq:sigma2R:Pzeta} gives rise to
\begin{equation}
    \sigma_R^2 = \frac{\calP_{\zeta}(k_\uend)}{50}\left\{5 - \frac{4}{3} \ln\left(\frac{H_\uend M}{4\pi}\right) \left[-3 - \frac{2}{3}\ln\left(\frac{H_\uend M}{4\pi}\right)\right] \right\} \left(\frac{H_\uend M}{4\pi}\right) ^ {-4/3}\, .
    \label{eq:sigmaR:appr}
\end{equation}
In this expression, we have used that the link between $R$ and $M$ is given by $M=4\pi\rho R^3/3 = 4\pi H^2 R^3 \Mp^2$. By plugging \Eqs{eq:deltac:constant} and~\eqref{eq:sigmaR:appr} into \Eq{eq:FPT:PS}, one thus obtains an explicit expression for the mass fraction $\beta(M)$, that we do not reproduce here since it is not particularly insightful, but which is nonetheless straightforward. 

Note that in general, the mass fraction has to be evolved from the time black holes form (which here depends on the mass) to the time at which $\beta$ is given, taking into account that PBHs may dilute at a different rate than the background energy density. However, here, given that the universe behaves as matter-dominated during the instability phase, $\beta$ remains constant so our result does correspond to the mass fraction at the time $t_\Gamma$, when the instability stops.

The corresponding formula is displayed in \Fig{fig:comoving} with the solid blue line, and one can check that it gives a reliable approximation to the full result.
It can thus be used to assess its overall integral, \ie $\OmegaPBH$, and the mass at which it peaks, which we now do.

By integrating \Eq{eq:FPT:PS} over $S$ (and reminding that $\delta_\uc$ does not depend on $S$), one has
\bea
\OmegaPBH=\erfc\left[\frac{\delta_\uc}{\sqrt{2 S_\umax}}\right],
\eea
where $S_\umax$ is the maximal value of $S$, \ie the one corresponding to the minimum value of $R$, or of $M$. It can thus be obtained by setting $M=4\pi\Mp^2/H_\uend$ in \Eq{eq:sigmaR:appr}, giving rise to $S_\umax=\calP_\zeta(k_\uend) / 10$. Making also use of \Eq{eq:deltac:constant}, one obtains
\begin{equation}
    \label{eq:Omegatot:anal}
\OmegaPBH = \erfc\left[\sqrt{\frac{5}{\calP_\zeta(k_\uend)}}\left(\frac{3\pi}{2} \frac{H_\Gamma}{H_\uend}\right) ^{2 / 3}\right]\, .
\end{equation}
This formula is displayed in the left panel of \Fig{fig:comoving} with the dashed blue line, and is found to provide a good fit to the full numerical result. Furthermore, it allows one to identify the region of parameter space in which the universe is dominated by PBHs at the end of the instability phase, $\OmegaPBH>1/2$, which reduces to
\bea 
\label{eq:bound:Omegatot_eq_1}
\frac{\rho_\Gamma^{1/4}}{\rho_\uend^{1/4}}< 8\times 10^{-6} \left(\frac{\rho_\uend^{1/4}}{10^{16}\mathrm{GeV}}\right)^{3/2}\, .
\eea 
This upper bound is displayed with the dashed blue line in the right panel of \Fig{fig:comoving}, and one can see that it indeed provides an accurate estimate of the boundary between the region in parameter space in which PBHs are very abundantly produced and the region in which they remain subdominant. Up to a prefactor of order one, this also matches Eq.~(4.1) of \Refa{Martin:2019nuw}. Indeed, the bound~\eqref{eq:bound:Omegatot_eq_1} corresponds to requiring that the instability phase be sufficiently long that when the linear theory is extrapolated throughout the oscillatory phase, the most amplified scales, \ie the ones around the Hubble radius at the end of inflation, reach a typical value for the density contrast, $\sqrt{\calP_\delta}$, that is of order one when the instability stops. This conclusion seems therefore to be robust to the inclusion of cloud-in-cloud effects, and to the detailed description of the mass distribution of PBHs, which the present, more refined analysis, allows for.

Our analytical approximation can also be used to estimate the typical mass at which the mass fraction peaks. From \Eq{eq:FPT:PS}, in the limit of a scale-invariant threshold, the distribution of first-crossing ``times'' peaks at $S_{\mathrm{peak}}=\delta_\uc^2/3$. In practice, this may select a mass that is smaller than the Hubble mass at the end of inflation, which then indicates that the mass fraction is a decreasing function of the mass, and is maximal near the Hubble mass at the end of inflation,
\bea
\label{eq:Mend}
M_\uend = \frac{4\pi\Mp^2}{H_\uend}\simeq 10\mathrm{g} \left(\frac{\rho_\uend^{1/4}}{10^{16}\mathrm{GeV}}\right)^{-2}\, .
\eea 
Otherwise, in the regime where $M\gg M_\uend$, \Eq{eq:sigmaR:appr} can be approximated by keeping the squared logarithmic term only, and in this limit one obtains
\bea
\label{eq:meanMass}
\frac{M_{\mathrm{peak}}}{\mathrm{g}}\simeq 1.22\times 10^{-9} \frac{\rho_\uend^{1/4}}{10^{16}\GeV} \left(\frac{\rho_\uend^{1/4}}{\rho_\Gamma^{1/4}}\right)^2
\ln^{3/2}\left[3.62\times 10^{-7}\left(\frac{\rho_\uend^{1/4}}{10^{16}\GeV}\right)^2\left(\frac{\rho_\uend^{1/4}}{\rho_\Gamma^{1/4}}\right)^{4/3}\right]\, .
\eea 
This value is displayed with the dashed green line in \Fig{fig:mass-function}, where one can check that it provides indeed a reliable estimate. It is interesting to notice that the condition~\eqref{eq:bound:Omegatot_eq_1} for an efficient production of PBHs is (roughly) equivalent to requiring that $M_\mathrm{peak}>M_\uend$. As a consequence, in the regime where PBHs are abundantly produced, the peak mass is substantially larger than the Hubble mass at the end of inflation, and hence corresponds to scales that emerge from the Hubble radius several \efolds~before the end of inflation. This is because, although smaller scales spend more time within the instability band and are thus more amplified, the initial value of their power spectrum is also smaller, and the trade-off selects intermediate scales.  
\section{Discussion and conclusion}
\label{sec:Discussion}
In this work, we have made use of the excursion-set approach to accurately compute the mass distribution of primordial black holes that are produced during metric preheating. The parametric instability of metric preheating occurs in any inflationary model in which the inflaton oscillates around a minimum of its potential after inflation. It is therefore a rather generic, not to say inevitable, phenomenon, for which it is thus important to precisely characterise the properties of the resulting black holes. Since metric preheating leads to the amplification of a wide range of scales, the cloud-in-cloud mechanism, in which small-mass black holes are trapped inside regions later collapsing into larger-mass black holes, plays an important role. This is why one needs to go beyond the simplified, Press--Schechter-inspired, common estimate to assess the mass distribution of black holes.

After reviewing the different techniques that have been proposed in the literature to compute mass distributions of PBHs, and highlighting salient aspects of these methods that are most of the time only alluded to, we have studied the problem at hand with the excursion-set formalism, combining different numerical techniques (namely a direct Monte-Carlo sampling of Langevin realisations and numerical solutions of Volterra integral equations) and analytical approximations.

Assuming that the potential of the inflaton is almost quadratic in the last stages of inflation (and during the oscillatory phase), the result only depends on two parameters, namely the energy density at the end of inflation, $\rho_\uend$, and the energy density at the time the inflaton decays into other degrees of freedom and the instability stops~\cite{Martin:2020fgl}, $\rho_\Gamma$. 

We have found that in the region of parameter space corresponding to \Eq{eq:bound:Omegatot_eq_1}, PBHs are very abundantly produced, in such a way that they even dominate the energy content of the universe at the end of the oscillatory phase. For this to happen, the inflaton needs to be sufficiently weakly coupled, such that more than 5 order of magnitude separate $\rho_\uend^{1/4}$ from  $\rho_\Gamma^{1/4}$, but as soon as this is the case, one is led to this rather drastic conclusion that the universe undergoes an early phase of PBH domination. 

The typical masses of the black holes range from 10 grams to the mass of the sun. For masses smaller than $10^9$ grams, PBHs evaporate before big-bang nucleosynthesis (BBN) and can therefore not be directly constrained. Heavier black holes would, however, survive until and after BBN, and given that the universe is radiation dominated at BBN, this excludes the region with $M>10^9$g in \Fig{fig:comoving-mean}. In fact, even if $M>10^9$g, it was recently shown in \Refa{Papanikolaou:2020qtd} that the gravitational potential underlain by a gas of PBHs induces the production of gravitational waves at seconder order in cosmological perturbation theory, and that these gravitational waves may lead to a backreaction problem if $\OmegaPBH>{10^{-4}}(10^9\mathrm{g}/M)^{1/4}$ at the time PBHs form. This excludes the value $\OmegaPBH\sim 1$ for any $M>10$g, hence the whole region described by \Eq{eq:bound:Omegatot_eq_1}, and located above the dashed blue line in \Fig{fig:comoving}, may be excluded too. 

This region was correctly identified in \Refa{Martin:2019nuw} already, in which metric preheating was studied with the simplified, Press--Schechter-inspired approach described in \Sec{sec:simplified:PS}. A detailed comparison between our result and the ones obtained in \Refa{Martin:2019nuw} is given in \App{sec:Comparison:Previous:Work}. The main difference we find concerns the mass at which the mass distributions peak. While in \Refa{Martin:2019nuw}, it was found to correspond to the Hubble mass at the end of inflation, see \Eq{eq:Mend}, in the present analysis we find that, in the regime where PBHs are substantially produced, the peak mass is substantially larger, see \Eq{eq:meanMass}. This is a consequence of the cloud-in-cloud mechanism, which could not be taken into account in \Refa{Martin:2019nuw}, and this has two main consequences. First, since heavier black holes take more time to Hawking evaporate, they survive for a longer period, hence the constraints on the parameters of the model arising from the present result are more stringent than the ones obtained in \Refa{Martin:2019nuw}, which made conservative assumptions as pointed out in that reference. Second, this implies that the bulk of the PBH population comes from modes that exit the Hubble radius several \efolds~before the end of inflation, at a stage where it is not clear that the potential can still be approximated as being quadratic. This means that, in practice, it may be necessary to analyse each potential individually instead of using the generic parametrisation employed in this work. We have, however, provided all the relevant formulae and technical considerations for such an exploration to be carried out (the numerical code we have developed to produce the results presented in this article is also publicly available in the arXiv ancillary files). 

Let us also highlight that, although a large range of scales is amplified during metric preheating, the mass distributions we have found are rather peaked, and never extend over more than a couple of orders of magnitude. This is in agreement with \Refa{MoradinezhadDizgah:2019wjf}, where the mass distribution associated to broad spectra was explored, and it was found to be quasi-monochromatic (\ie peaked at a single mass), which corresponds to either the smallest enhanced scale or the largest enhanced scale, depending on the tilt of the spectrum. In the present case, we found that the peak mass arises at rather intermediate scales, but this is because the cloud-in-cloud mechanism plays an important role (such that the mass fraction at small masses is suppressed), while \Refa{MoradinezhadDizgah:2019wjf} focused on regimes where the mass fraction remains small and the cloud-in-cloud phenomenon is almost absent. 

Another remark of interest is that, as argued in \Sec{sec:Remove:Super:Horizon}, we have evaluated the density contrast in comoving slices in order to apply our PBH formation criterion. In \App{sec:NewtonGauge}, we investigate the consequences on interpreting the formation threshold in a different slicing (namely the Newtonian one). While we find that most conclusions are unchanged, the main difference is that the mass distributions are much wider when working in the Newtonian slicing. This is because, in the Newtonian gauge, the density contrast is not suppressed on super-Hubble scales, hence large-scale fluctuations substantially contribute to the coarse-grained density perturbation inside a Hubble patch. This is precisely the effect we have tried to avoid by using the comoving slicing, since large-scale fluctuations should only lead to a local rescaling of the background, and not determine the fate of overdensities inside the Hubble radius~\cite{Young:2014ana}.

Let us also note that our analysis was restricted to scales that are larger than the Hubble radius at the end of inflation, and we did not explicitly compute the mass fraction at smaller scales. However, there is a small range of scales that are within the Hubble radius at the end of inflation, but that still enter from below the instability band during the oscillatory phase (see the lowest dotted line in \Fig{fig:scale}). Although the physical status of those scales is unclear (since they remain within the Hubble radius throughout inflation, they behave as Minkowski vacuum fluctuations, and never undergo classical amplification~\cite{Polarski:1995jg, Lesgourgues:1996jc, Kiefer:2008ku, Martin:2015qta}), they can nonetheless be readily incorporated in the excursion-set approach. When doing so, given that the corresponding range of scales is very narrow, and that the initial density contrast is tiny (since those scales are not excited during inflation), we find that this only adds a negligible, low-mass end to the mass distributions we have computed, so these scales can be safely discarded.  

A final remark of interest is that the collapse criterion we have employed was derived in \Refa{Goncalves:2000nz} for a universe filled with a scalar field with quadratic potential by assuming a spherically symmetric profile for the overdensity. In the case of a universe filled with a pressureless perfect fluid, it is well known that PBH peaks no longer need to be rare and hence may not be close to spherically symmetric, and that corrections arising from spherical asymmetries typically lead to less abundant PBHs (see for instance \Refa{Harada:2016mhb}). Although those two systems are different, as discussed in detail in \Refa{Martin:2020fgl}, one may expect that a similar effect takes place in the setup under consideration in this work. In order to address it, one would have to generalise the calculation of \Refa{Goncalves:2000nz} to non-spherical geometries, which could be the topic of future work.

In summary, by properly taking into account the cloud-in-cloud mechanism, which plays an important role in the metric preheating instability in which a large range of scales is enhanced, we have derived accurate predictions for the mass distribution of primordial black holes produced during preheating. Given that those black holes may dominate the universe for a transient period afterwards, reheat the universe by Hawking evaporation, and induce a detectable stochastic gravitational wave background~\cite{Papanikolaou:2020qtd}, the details of these mass distributions may indeed have important cosmological consequences, some of which remain to be explored.
\begin{acknowledgments}
It is a pleasure to thank Karsten Jedamzik, J\'er\^ome Martin, Theodoros Papanikolaou and Dani\`ele Steer for interesting discussions. PA thanks Gilles Chabrier without whom this work would probably not have seen the light of day. 
\end{acknowledgments}

\appendix
\section{Numerical solution of the Volterra equation}
\label{sec:numerical-volterra}
In \Sec{sec:Volterra} we have shown that the first-crossing-time distribution associated to the Langevin equation~\eqref{eq:Langevin:sigma2} satisfies a family of Volterra integral equation, one of them being of the form
\begin{equation}
	\begin{aligned}
		\Pfpt(S) =& \left[\frac{\delta_\uc(S)}{S}-\delta_\uc'(S)\right] P_\mathrm{free}\left[\delta_\uc(S),S\right]
		\\ &
		+\int_0^S \dd s \left[\delta_\uc'(S)- \frac{\delta_\uc(S)-\delta_\uc(s)}{S-s}\right]P_\mathrm{free}\left[\delta_\uc(S)-\delta_\uc(s),S-s\right] \Pfpt(s)\, ,
	\end{aligned}
\end{equation}
see \Eq{eq:Volterra:regular}. Upon discretising the time variable according to $S = n \Delta s$ and $s = m \Delta s$, where $n$ and $m$ are integer numbers and $\Delta S$ is a numerical time step, the Volterra equation can be written as
\begin{equation}
	\Pfpt^n = X^n + \tensor{M}{^n_m} \Pfpt^m\, ,
	\label{eq:matrix:product}
\end{equation}
where the implicit summation notation is employed. In this formula, $\Pfpt^n$ and $X^n$ are vectors, and $\tensor{M}{^n_m}$ is a lower triangular matrix, with non-zero cells only if $m \leq n $, defined as
\begin{align}
    \Pfpt^n & = \Pfpt(n\Delta S)\\
	X^n &= \left[\frac{\delta_\uc(n \Delta s)}{n \Delta s}-\delta_\uc'(n\Delta s)\right] P_\mathrm{free}\left[\delta_\uc(n\Delta s),n\Delta s\right]\\
	\tensor{M}{^n_m} &= \left[\delta_\uc'(n\Delta s)- \frac{\delta_\uc(n\Delta s)-\delta_\uc(m\Delta s)}{(n-m) \Delta s}\right]P_\mathrm{free}\left[\delta_\uc(n\Delta s)-\delta_\uc(m\Delta s), (n-m) \Delta s\right].
\end{align}
Note that we chose to solve \emph{the} Volterra equation that is such that the diagonal of the matrix $\tensor{M}{^n_m} $ is $0$.
If one tries to discretise \Eq{eq:Volterra:kernel} in general, then one obtains a matrix $\tensor{M}{^n_m}$ with diverging elements on the diagonal, unless the specific choice~\eqref{eq:Volterra:regular} is made, which proves its usefulness.

The solution to \Eq{eq:matrix:product} is simply given by
\begin{equation}
	\Pfpt = (\mathrm{Id} - M )^{-1} X.
\end{equation}
Since $\mathrm{Id} - M $ is a lower triangular matrix, it can be easily inverted with the ``forward-substitution'' algorithm, the numerical cost of which only scales as the square of the size of the matrix (compared to cubic scaling in general).
\section{Density contrast in the Newtonian slicing}
\label{sec:NewtonGauge}
\begin{figure}
\centering
\begin{subfigure}{0.445\textwidth}
	\includegraphics[width=\textwidth]{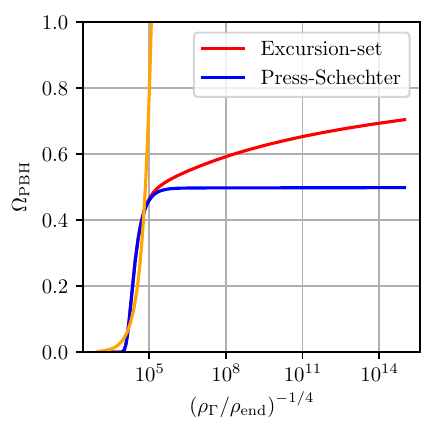}
	\caption{Evolution of $\OmegaPBH$ at $\rho_\uend = 10^{-12}  \Mp^{4}$}
	\label{fig:newton-omega}
\end{subfigure}
\begin{subfigure}{0.545\textwidth}
	\includegraphics[width=\textwidth]{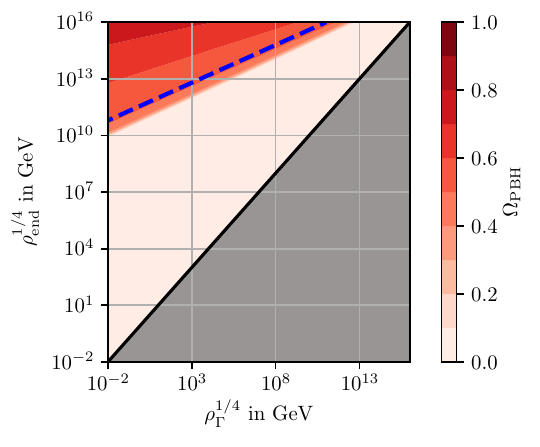}
	\caption{$\OmegaPBH$ as a function of $\rho_\uend$ and $\rho_\Gamma$}
	\label{fig:newton-omega-paramSpace}
\end{subfigure}
\begin{subfigure}{0.496\textwidth}
	\includegraphics[width=\textwidth]{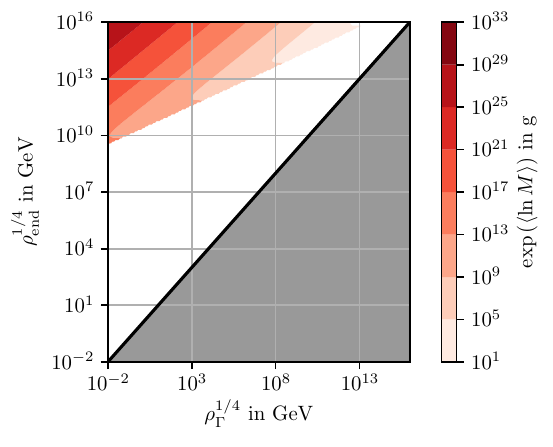}
	\caption{Average mass of PBH}
	\label{fig:newton-mean}
\end{subfigure}
\begin{subfigure}{0.496\textwidth}
	\includegraphics[width=\textwidth]{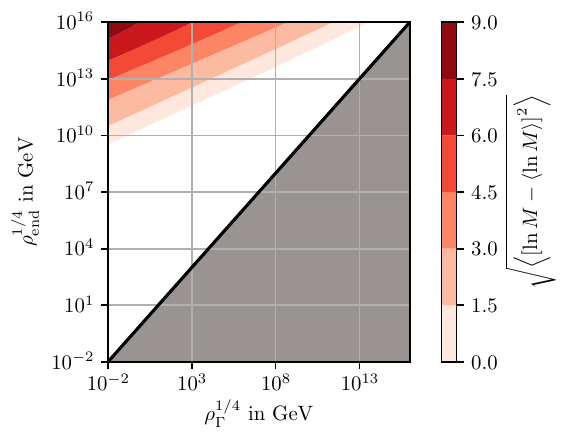}
	\caption{Dispersion of the PBH masses}
	\label{fig:newton-dispersion}
\end{subfigure}
\caption{PBH mass fraction if the formation criterion is interpreted in the Newtonian slicing.}
\label{fig:newton}
\end{figure}
In \Sec{sec:Remove:Super:Horizon}, we motivated the choice of the comoving slicing to evaluate the density contrast on super-Hubble scales, and express our PBH formation criterion.
In the present appendix, we explore the consequences of choosing a different slicing, in order to understand how much our conclusions depend on that choice. In practice, we consider the Newtonian slicing, which consists in setting $\alpha=\beta=-2$ in \Eq{eq:delta:Phi}, and \Eq{eq:delta:Phi:MD} reduces to
\begin{equation}
	\delta_k \simeq  -\frac{2}{5} \left[3 + \left(\frac{k}{aH}\right)^2\right]\zeta_k \, .
\end{equation}
The main difference with the comoving slicing is that, for a curvature perturbation that is scale invariant on super-Hubble scales, the density contrast is not suppressed anymore above the Hubble radius. This is why we discarded this choice of slicing in \Sec{sec:Remove:Super:Horizon}, since super-Hubble scales should only lead to a local rescaling of the background field values inside a Hubble patch, and thus not contribute to whether or not a black hole forms. 

In the Newtonian slicing, on super-Hubble scales, $\delta_k \simeq  -\frac{6}{5}\zeta_k$, so $\delta_k$ is conserved since $\zeta_k$ is (and contrary to the comoving slicing where $\delta_k$ grows like the scale factor). This means that the variance of the density contrast can equally be evaluated at the end of inflation or at the band-crossing time, namely
\begin{equation}
	\sigma_R^2(t_\uend) = \sigma_R^2(t_\mathrm{bc}) =  \left(\frac{6}{5}\right)^2\int_0^{a/R} \calP_\zeta\left(k,t_\uend\right) \frac{\dd k}{k}\, .
\end{equation}
The collapse criterion is given by \Eq{eq:deltac:bc}, and it is strongly scale dependent, contrary to the case of the comoving density contrast.
As explained in section \ref{sec:RedThreshold}, a good approximation in the case of very red thresholds is given by the Press-Schechter formula without the additional factor of $2$.
This is explained by the fact that the barrier moves faster than the average trajectories and multiple crossings are unlikely.

We apply the techniques described in \App{sec:numerical-volterra} and present our results in \Fig{fig:newton}.
In \Fig{fig:newton-omega}, one can check that indeed, the (one times) Press-Schechter formula provides a good approximation to the excursion-set result, up until $\OmegaPBH$ reaches $1/2$.
This is because the Press-Schechter formula does not allow for more than half of the universe being collapsed.
One can check that the condition~\eqref{eq:RedCriterion} is indeed verified only when $\OmegaPBH < 1/2$.
For that purpose we set $\epsilon \approx 1/2$ and, in \Fig{fig:newton-omega}, we display
\begin{equation}
    \max_{\ln(R_1/R_2) > \epsilon}\left[\frac{\sqrt{S(R_2) - S(R_1)}}{\delta_\uc(R_1) - \delta_\uc(R_2)}\right]
\end{equation}
with the solid yellow line.
One expects the Press-Schechter formula to provide a good approximation when this quantity is below unity, which is indeed the case. Above $\OmegaPBH=1/2$, the abundance of PBHs continues to slowly increase.
This behaviour is rather different from the results obtained with the comoving density contrast in \Sec{sec:NumericalResults}, where we found an abrupt transition from $\OmegaPBH=0$ to $\OmegaPBH = 1$, see \Fig{fig:comoving}.

Beside those differences, by comparing the right panel of \Fig{fig:comoving} with \Fig{fig:newton-omega-paramSpace}, one can see that the region of parameter space that substantially produces PBHs is roughly the same (in \Fig{fig:newton-omega-paramSpace}, we have reported the dashed blue line of the right panel of \Fig{fig:comoving} in order to guide the reader's eye). By comparing \Fig{fig:comoving-mean} with  \Fig{fig:newton-mean}, one can see that the typical masses involved are also roughly the same. However, when comparing \Figs{fig:comoving-dispersion} and~\ref{fig:newton-dispersion}, one realises that the mass distributions are much wider in the Newtonian case. This is because, since the density contrast is not suppressed on large scales in the Newtonian slicing, it yields a heavier large-mass tail than in the comoving slicing.  
\section{Comparison with \Refa{Martin:2019nuw}}
\label{sec:Comparison:Previous:Work}
\begin{figure}
    \centering
    \includegraphics{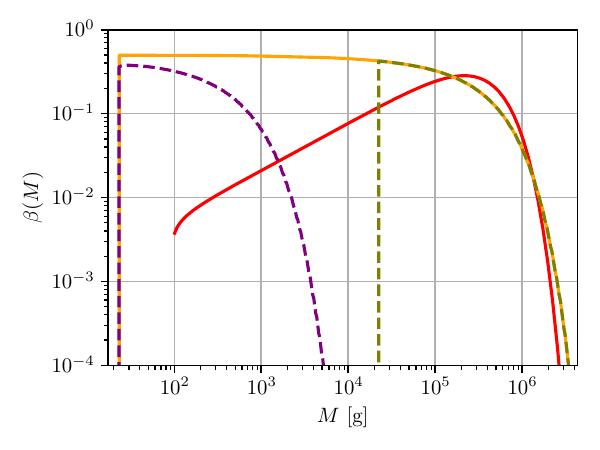}
    \caption{Mass fraction $\beta$ of primordial black holes for $\rho_\uend=10^{-12}\Mp^4$ and $\rho_\Gamma=10^{-40}\Mp$. As in \Fig{fig:mass-function}, the red line corresponds to numerically solving the Volterra equation~\eqref{eq:Volterra:regular}. The olive, orange and purple lines correspond to the results of \Refa{Martin:2019nuw} and are taken from Fig.~4 of that reference. The orange line displays the ``raw'' result obtained with the estimate of \Sec{sec:simplified:PS}, which leads to the problematic $\OmegaPBH>1$. Then, ``renormalisation'' is either performed by ``premature ending'' (purple dashed line) or by ``absorption'' (olive dashed line).}
    \label{fig:PreviousWorks}
\end{figure}
In \Refa{Martin:2019nuw}, a calculation of the PBH abundance in metric preheating was performed by making use of the simplified, commonly used estimate presented in \Sec{sec:simplified:PS}. While this is enough to correctly identify the region in parameter space that leads to substantial PBH production, see the discussion below \Eq{eq:bound:Omegatot_eq_1}, which is consistent with Eq.~(4.1) of \Refa{Martin:2019nuw}, this is a priori not sufficiently accurate to derive detailed predictions about the mass distribution, in particular in regimes where the cloud-in-cloud mechanism plays an important role.

In \Refa{Martin:2019nuw}, it was indeed pointed out that in cases where PBHs are abundantly produced, \ie under the condition~\eqref{eq:bound:Omegatot_eq_1}, the simple estimate of \Sec{sec:simplified:PS} predicts $\OmegaPBH>1$, which clearly signals its breaking down. In order to deal with this issue, two solutions were proposed in \Refa{Martin:2019nuw}: either remove by hand the small-mass end of the distribution, in order to bring $\OmegaPBH$ back to one, and to model the possible absorption of small black holes into larger black holes (this was dubbed ``renormalisation by absorption''); or stop the instability phase prematurely, at the time when $\OmegaPBH$ crosses one, since the universe stops being dominated by an oscillating scalar field at that point (this was dubbed ``renormalisation by premature ending''). Since the later approach effectively removes larger-mass black holes from the distributions (given that heavier black holes come from larger scales, that enter the instability band later), it was assumed in \Refa{Martin:2019nuw} that these two results would bound the true mass distribution on each side, and that any conclusion that can be drawn in both approaches probably applies to the actual result. 

In this appendix, we want to verify these statements, and in \Fig{fig:PreviousWorks}, we compare the mass fraction obtained in this work (solid red line) with the formulas derived in \Refa{Martin:2019nuw} when ``renormalisation'' is performed by absorption (olive line) or by premature ending (purple line), starting from the ``raw'' result (orange line) that leads to the problematic $\OmegaPBH>1$. Although, as expected, neither approach provides a good description of the full result, the order of magnitude of the overall amplitude is correctly reproduced, and the actual mass distribution is indeed approximately bounded by the results from the two renormalisation procedures (although it is closer to the ``renormalisation by absorption'' result). This therefore confirms the relevance of \Refa{Martin:2019nuw}. 

The main difference in the shape of the mass distribution concerns the location of the peak mass: in  \Refa{Martin:2019nuw}, it was found that the mass where most PBHs form is the smallest mass that undergoes parametric amplification, \ie the Hubble mass at the end of inflation, see \Eq{eq:Mend}, or the cutoff mass in the case of ``renormalisation by absorption''. Here, we find that the mass distribution peaks several orders of magnitude above that mass, see the discussion following \Eq{eq:meanMass}. %
\bibliographystyle{JHEP}
\bibliography{excursion_set}
\end{document}

%% file: newcommands.tex
\newcommand{\ie}{{i.e.~}}

\newcommand{\eg}{e.g.~}

\newcommand{\apriori}{{a priori~}}
\newcommand{\etc}{\textsl{etc.~}}

\newcommand{\np}{n_\mathrm{points}}

\newcommand{\Dirac}{\delta_{\sss{\mathrm{D}}}}
\newcommand{\Pfpt}{P_{\sss{\mathrm{FPT}}}}

\let\oldsqrt\sqrt
\def\sqrt{\mathpalette\DHLhksqrt}
\def\DHLhksqrt#1#2{%
\setbox0=\hbox{$#1\oldsqrt{#2\,}$}\dimen0=\ht0
\advance\dimen0-0.2\ht0
\setbox2=\hbox{\vrule height\ht0 depth -\dimen0}%
{\box0\lower0.4pt\box2}}

\newcommand{\order}[1]{\mathcal{O}\!\left(#1\right)}

\DeclareMathOperator{\erfc}{erfc}

\newcommand{\dd}{\mathrm{d}}
\newcommand{\ee}{e}

\newcommand{\sss}[1]{{\scriptscriptstyle{#1}}}

\newcommand{\uPl}{\mathrm{Pl}}
\newcommand{\uin}{\mathrm{in}}

\newcommand{\umax}{\mathrm{max}}
\newcommand{\uend}{\mathrm{end}}

\newcommand{\uc}{\mathrm{c}}

\newcommand{\usssPl}{\sss{\uPl}}

\newcommand{\calP}{\mathcal{P}}

\newcommand{\GeV}{\mathrm{GeV}}

\newcommand{\Mp}{M_\usssPl}

\newcommand{\efolds}{$e$-folds}

\newcommand{\beq}{\begin{equation}}
\newcommand{\eeq}{\end{equation}}
\newcommand{\bea}{\begin{equation}\begin{aligned}}
\newcommand{\eea}{\end{aligned}\end{equation}}

\newlength{\wsingfig}
\newlength{\wdblefig}
\newlength{\wquadfig}
\newlength{\wtriplefig}
\newcommand{\Eq}[1]{Eq.~(\ref{#1})}
\newcommand{\Eqs}[1]{Eqs.~(\ref{#1})}
\newcommand{\Fig}[1]{Fig.~{\ref{#1}}}
\newcommand{\Figs}[1]{Figs.~{\ref{#1}}}
\newcommand{\Refa}[1]{Ref.~{\cite{#1}}}
\newcommand{\Refs}[1]{Refs.~{\cite{#1}}}
\newcommand{\Sec}[1]{Sec.~\ref{#1}}
\newcommand{\Secs}[1]{Secs.~\ref{#1}}
\newcommand{\App}[1]{Appendix~\ref{#1}}